\documentclass[twocolumn,showpacs,prb,superscriptaddress]{revtex4}

\usepackage{graphicx}

\begin{document}

\title{Measuring the spin polarization and Zeeman energy of a spin-polarized electron gas: Comparison between Raman 
scattering and photoluminescence}
\author{Cynthia Aku-Leh}
\email{cynthia.aku-leh@kcl.ac.uk}
\affiliation{Institut des Nanosciences de Paris, UMR 7588, CNRS/Universit\'e Paris VI et VII, Campus Boucicaut, 140 rue 
de Lourmel, 75015 Paris, France}
\affiliation{Department of Physics, King's College London, Strand, London WC2R 2LS, United Kingdom}
\author{Florent Perez}
 \email{Florent.Perez@insp.jussieu.fr}
\affiliation{Institut des Nanosciences de Paris, UMR 7588, CNRS/Universit\'e Paris VI et VII, Campus Boucicaut, 140 rue 
de Lourmel, 75015 Paris, France}
\author{Bernard Jusserand}
\affiliation{Institut des Nanosciences de Paris, UMR 7588, CNRS/Universit\'e Paris VI et VII, Campus Boucicaut, 140 rue 
de Lourmel, 75015 Paris, France}
\author{David Richards}
\affiliation{Department of Physics, King's College London, Strand, London WC2R 2LS, United Kingdom}
\author{Wojciech Pacuski}
\affiliation{Institute of Experimental Physics, Warsaw University, Ho\.za 69, PL-00-681 Warszawa, Poland}
\affiliation{Institut N\'eel/CNRS-Universit\'e J. Fourier, Bo\^{\i}te Postale 166, F-38042 Grenoble Cedex 9, France}
\author{Piotr Kossacki}
\affiliation{Institute of Experimental Physics, Warsaw University, Ho\.za 69, PL-00-681 Warszawa, Poland}
\author{Michel Menant}
\affiliation{Institut des Nanosciences de Paris, UMR 7588, CNRS/Universit\'e Paris VI et VII, Campus Boucicaut, 140 rue 
de Lourmel, 75015 Paris, France}
\author{Grzegorz Karczewski}
\affiliation{Institute of Physics, Polish Academy of Sciences, al. Lotnik\'ow 32/46, 02-668 Warszawa, Poland}

\date{\today}
\begin{abstract}
We compare resonant electronic Raman scattering and photoluminescence measurements for the characterization of a 
spin-polarized two-dimensional electron gas embedded in $\text{Cd}_{1-x}\text{Mn}_x\text{Te}$ single quantum wells. 
From Raman scattering by single-particle excitations in a zero magnetic field, we measure the Fermi velocity and then 
obtain the Fermi energy (as well as the electron density), which is comparable to that extracted from  
photoluminescence for moderate electron densities, assuming a bare band-edge mass. At large electron densities, the 
Fermi energies derived from Raman scattering and photoluminescence differ. For an applied in-plane magnetic field and 
zero wave vector transferred to the electron gas, Raman scattering spectra show peaks at both the Zeeman energy $Z$, 
resulting from collective excitations of the spin-polarized electron gas, and the one electron spin-flip energy $Z^*$. 
Magneto-photoluminescence spectra show conduction band splitting that are equivalent to $Z$, suggesting that 
collective effects are present in the photoluminescence spectra. Assuming (as before) an uncorrected mass, the degree 
of spin polarization $\zeta$ determined from the magneto-photoluminescence lineshape is found to differ from that 
derived from the magnetic field dependent Raman scattering measurements for large electron densities. We attribute the 
discrepancy in measuring $\zeta$ and the Fermi energy to the renormalized mass resulting from many-body 
electron-electron interactions. 

\end{abstract}


\maketitle

\section{Introduction}

Over the last couple of decades, two-dimensional electron gases (2DEGs) embedded in quantum wells have provided a 
unique means for understanding many-body exchange and correlation effects (of the Coulomb interaction). However, 
direct study of spin-polarized two-dimensional electron gases, important for the understanding of spin-physics, only 
began in the last decade. This is due in part to advances in growth techniques of dilute semi-magnetic quantum wells 
with high electron mobilities,\cite{karczewski} especially those of II-VI materials which serve as ideal systems to 
study the spin-polarized case, and in part to applications in spintronics -- the use of electron spin in semiconductor 
devices as opposed to electron charge.\cite{Awschalom1} 

In II-VI paramagnetic heterostructures, spin polarization of the 2DEG is achieved through the application of an 
external magnetic field. This produces a large Zeeman splitting, induced through exchange interaction between the 
conduction band electrons of the quantum well and the Mn$^{2+}$ ions.\cite{gaj} This means that in small magnetic 
fields, spin quantization dominates over orbital quantization leading to a spin-polarized 2DEG rather than quantum 
Hall states as found in GaAs quantum wells.\cite{Pinczuk2, Pinczuk} Further, for the same electron density $n_s$, the 
dimensionless coupling constant $r_s$, which describes the strength of many-body electron-electron interactions, 
defined as $1/(a_{\text{B}}\sqrt{\pi n_s})$, is large due to the small Bohr radius $a_{\text{B}}$ that these materials 
exhibit. Corrections due to  exchange-correlations, therefore, become important and will inevitably renormalize the 
electron mass\cite{Tan1, Smith, Asgari} and the Zeeman splitting,\cite{Jusserand4} and hence, affect the manner in 
which the spin polarization degree and the Fermi energy of an electron gas are determined by spectroscopy. 

In this work, we present a quantitative comparative study of fundamental parameters -- the Zeeman energies, the spin 
polarization degree, the electron density and the Fermi energy -- of a spin-polarized 2DEG embedded in Cd$_{1-x}$Mn$_x$
Te single quantum wells using resonant electronic Raman scattering and photoluminescence (PL) in both zero and applied 
magnetic fields in the Voigt configuration. We will demonstrate that our PL results show collective behavior, allowing 
us to extract the bare Zeeman energy instead of the renormalized one while Raman scattering measurements show both. 
Additionally, we will show that for the range of densities accessible by us ($r_s$ between 1.8 and 3), an 
understanding of the renormalization of mass due to exchange-correlation effects is needed to explain the difference 
we observe in our measurements for the spin polarization and the Fermi energy for large electron densities 
using both Raman scattering and PL. 

PL is a widely used spectroscopic tool for probing inter-band electronic excitations.\cite{Peric, Bernard4, Kossacki, 
Masalana, Hawrylak, Skolnick1, Teran, Hourd} In a zero magnetic field, PL corresponds to electrons excited by 
absorption of photons above the absorption edge of the 2DEG: an electron is excited into the conduction band and a 
hole is created in the valence band. The hole then relaxes to the top of the valence band where it becomes localized 
due to ionized impurities and potential fluctuations, while electrons thermalize with the Fermi sea. This is followed 
by a recombination of the conduction band electrons with the localized holes. Wave vector conservation is relaxed, 
allowing for recombination of all electrons with any finite wave vector. The result is a broad lineshape, the spectral 
width of which corresponds to the Fermi energy (neglecting Coulomb interaction of the recombining electron and hole). 
In an applied magnetic field, both electron and hole bands split into two spin subbands, and recombination processes 
occur between these bands (see section IV.B.2).

Raman scattering is also a well-established tool for studying low energy intra-band elementary excitations of electron 
gases embedded in semiconductors. In the polarized configuration, in which the incoming and the outgoing light have 
the same polarization, Raman scattering probes charge-density fluctuations (plasmons) and single-particle excitations 
of an electron gas.\cite{Jusserand1, Jusserand2, Pinczuk, Pinczuk2, Pinczuk3, Fasol} In the depolarized configuration, 
in which the incoming and outgoing light polarization are orthogonal to each other, it probes, in zero applied 
magnetic field, spin-density excitations\cite{Pinczuk} and single-particle excitations,\cite{Jusserand3} and 
in an applied magnetic field, it also probes spin-flip excitations.\cite{Pinczuk4, Jusserand4, Sarma, Perez1}

This paper is arranged as follows. We describe and define the various excitations of a spin polarized two-dimensional 
electron gas, the bare and the renormalized Zeeman splitting energies, and the spin polarization in sections II.A and 
II.B. In section II.C we compare a few methods for obtaining the Fermi energy and the carrier density. The samples and 
experimental setup are presented in section III. Experimental results are given in section IV for both Raman 
scattering and PL measurements in two parts: section IV.A presents the zero magnetic field results for the 
determination of the Fermi energy and section IV.B presents the magnetic field dependent measurements for deducing 
the degree of spin polarization, the bare and renormalized Zeeman splitting, $Z$ and $Z^*$, respectively. We 
will discuss our results in-depth in section V, beginning with a comparison of the obtained Fermi energy values in 
section V.A; an interpretation of the PL lineshape and the Raman scattering results in terms of collective and 
single-particle behavior is given in section V.B, and in section V.C we end with a discussion of the spin 
polarization values obtained from PL and Raman scattering. Finally, we conclude in section VI.  

\section {Spin-polarized two-dimensional electron gas and the carrier density}
\subsection {Single-particle excitations and the Zeeman energies}

Single-particle excitations (SPE) correspond to the kinetic energy change due to the transfer of wave vector to an 
electron gas by an exciting light source. To access the true excitation spectrum of single electrons across the Fermi 
disk, Raman scattering in the polarized configuration is used since screening mechanisms due to plasmons are killed 
under strong resonance conditions.\cite{Sarma} Thus, in a zero magnetic field, the dependence of the SPE spectra on 
the in-plane wave vector gives the Fermi velocity. From the Fermi velocity and independent knowledge of the effective 
mass,\cite{cynt3} the electron density and the Fermi energy can be deduced for an electron gas (see section IV.A). 

In an applied magnetic field and for zero wave vector transferred to the electron gas, the energy of the collective 
spin-flip excitation, corresponding to all electrons simultaneously flipping their spin from spin-down to spin-up and 
vice versa, has been shown to follow Larmor$'$s theorem\cite{Landau} and to equal the bare Zeeman energy $Z$ [see 
insert (a) of Fig.~\ref{fig:MagnetoRam}].\cite{Jusserand4, Perez1} In the presence of electron-electron interactions, 
the energy of the single electron spin-flip excitation, corresponding to a single electron flipping its spin, is 
shifted from $Z$ by a local exchange-correlation field and leads to a renormalized Zeeman energy $Z^*$ [see insert (b) 
of Fig.~\ref{fig:MagnetoRam}].\cite{Jusserand4, Perez1, Florent2}

\subsection{Spin polarization}

The degree of spin polarization of an electron gas is defined as 
\begin{equation}
\zeta = (n_\uparrow-n_\downarrow)/(n_\uparrow + n_\downarrow) = (k_{\text{F}\uparrow}^2- k_{\text{F}\downarrow
}^2)/(k_{\text{F}\uparrow}^2+{k_{\text{F}\downarrow}^2}) \label{Zeta}, %
\end{equation}
where $n_\uparrow$ ($n_\downarrow$) is the density of spin-up (spin-down) electrons, and $k_{\text{F}\uparrow}$
($k_{\text{F}\downarrow}$) is the Fermi wave vector for the spin-up (spin-down) electrons. Assuming the same effective 
mass for the spin-up and spin-down subbands, which are split by $Z^*$ [refer to the inserts of 
Fig.~\ref{fig:MagnetoRam}], $\zeta$ can also be expressed as: 
\begin{equation}
\zeta = -Z^{*} /2 E_{\text{F}}(0),\label{zeta}
\end{equation}
where $E_{\text{F}}(0)$ [= $\hbar^2 \pi n_s/m^*$, $m^*$ is the effective mass] is the Fermi energy in zero magnetic 
field. 

A few groups have considered the spin polarization degree of 2DEGs in quantum wells. For example, in a previous study, 
Lema\^itre and co-authors pointed out the full spin polarization state of electrons embedded in Cd$_{0.02}$Mn$_{0.98}$Te quantum wells using magneto-absorption studies in the Faraday geometry.\cite {Lemaître} 
Independently, Astakhov $et$ $al.$\cite{Astakhov} also deduced the spin polarization degree from the oscillator 
strength of charged excitons in CdMnTe quantum wells using magneto-reflectivity measurements. In the first case, the 
spin-polarized physics is dominated by Landau quantization effects in the Faraday configuration. In the second case, 
low electron densities were considered with trion and exciton states dominating. For a degenerate electron gas out 
of Landau quantization, it is important to find alternative means of obtaining the Coulomb modified spin 
polarization. We will show that this possibility is afforded by both Raman scattering and PL. 

\subsection{Density and Fermi energy measurements}

Carrier density, and essentially the Fermi energy by the definition above, in modulation-doped quantum wells have been 
measured by several means in the past. We discuss a few of these methods as follows. For low ($\sim10^{10} \text{ 
cm}^{-2}$) to very low electron densities ($\sim10^9 \text{ cm}^{-2}$), optical detection of (dimensional) 
magneto-plasma resonance was used to determine electron densities in III-V quantum wells,\cite{Gubarev} and for high 
mobility carriers and concentrations, magneto-transport methods have been used for III-V materials.\cite{Kukushkin} 
However, due to low mobility carriers in II-VI systems as compared to III-V systems, it is difficult to measure the 
carrier density by these means since the cyclotron frequency is smaller than the inverse of the relaxation time to be 
observed by far-infrared or microwave spectroscopy. We note that although our samples have high carrier mobilities and 
concentrations, the presence of Mn impurities renders transport measurements difficult. For such systems, carrier 
concentrations have been determined by other means: for example, by measuring the Moss-Burstein shift, in which the 
difference between the PL maximum and the absorption peak is determined; \cite{Kossacki, Teran} by filling factors of 
Landau levels in transmission,\cite{Kossacki, Hourd} and through magneto-reflectivity spectra of charged excitons.\cite
{Astakhov}

An additional way of extracting carrier concentration, which has worked for II-VI and III-V systems, is the 
measurement of the PL linewidth;\cite{Shields, Skolnick1} however, due to disorder effects and wave vector break 
down, this method can sometimes provide inaccurate information (as we will show later in section IV. A. 2). An 
adequate fitting model is necessary to determine the density from the PL lineshape.\cite {ChristenTheory} Another 
method for extracting carrier densities in quantum wells is dispersive Raman scattering of intra-subband plasmons 
which has been demonstrated for both III-V systems \cite{DavidPlasmons} as well as II-VI systems. \cite{Jusserand1} 
Dispersive Raman scattering by plasmons, while shown to provide an excellent estimation of the electron 
density, also suffers from drawbacks. The observation of the plasmon mode in certain materials is limited by disorder 
and it becomes  difficult to access by Raman scattering.\cite{cynt2} In this paper, we will focus on the determination 
of the electron density and the Fermi energy from an analysis of the PL lineshape and by Raman scattering of 
single-particle excitations. With the exception of a few reports,\cite{Tan1, Matsuda, SmithStiles, Tang} the above 
means of measuring the density (the Fermi energy) have not considered the influence of the mass renormalization due to 
many-electron interaction, which we will show is important. 

\section{Samples and experimental setup}

Raman scattering and PL measurements were carried out on several samples with differing Mn$^{2+}$ concentration and 
electron density. The structures were grown by molecular beam epitaxy on GaAs substrates.\cite{karczewski} The Mn$^{2+}$ concentration ranged from 0.46 \% to 1.1 \% and was determined from Raman scattering measurements (see section 
IV. B.1). As a representative, one sample that we will constantly refer to for our comparative study, sample B, had a 
15 nm thick $\text{Cd}_{1-x}\text{Mn}_x\text{Te}$ quantum well with Mn$^{2+}$ concentration of $x = 0.75 \%$. The 
barriers of our quantum wells were made of $\text{Cd}_{1-y}\text{Mg}_y\text{Te}$ with $y=$ 15 \% Mg. Modulation doping 
was achieved by introducing iodine within the $\text{Cd}_{1-y}\text{Mg}_y\text{Te}$ top barrier with a spacer 
thickness of 40 nm. Our samples were immersed in superfluid helium ($\sim 1.5$ K). A tunable Ti-sapphire laser (pumped 
by an Ar$^+$ laser), with power densities below 0.1 W/cm$^2$ to avoid heating the Mn$^{2+}$ ions, was used as exciting 
laser source. The laser was tuned to resonate close to the transitions between the first conduction band and the first 
heavy-hole band of the quantum well ($E_1H_1$ absorption edge) at $\approx$ 1.62 eV. For magnetic field dependent 
studies, our samples were mounted in the center of a 4.5 T superconducting solenoid producing a magnetic field in the 
plane of (Voigt configuration) and perpendicular to (Faraday configuration) the quantum well. Raman scattering 
measurements were carried out in the back-scattering geometry.

\section{Experimental results}
\subsection{Zero magnetic field}

In this section, we will determine the Fermi energy and the electron density of our 2DEG systems using both Raman 
scattering and PL. We will begin with the determination of the Fermi velocity and then the Fermi energy 
$E_{\text{F, Raman}}(0)$ by dispersive Raman scattering of SPEs. Later, we present results on the PL lineshape and 
deduce the Fermi energy $E_{\text{F, PL}}(0)$.

\subsubsection{Fermi velocity determination from Raman scattering by SPEs}

\begin{figure}
\includegraphics[width=\linewidth]{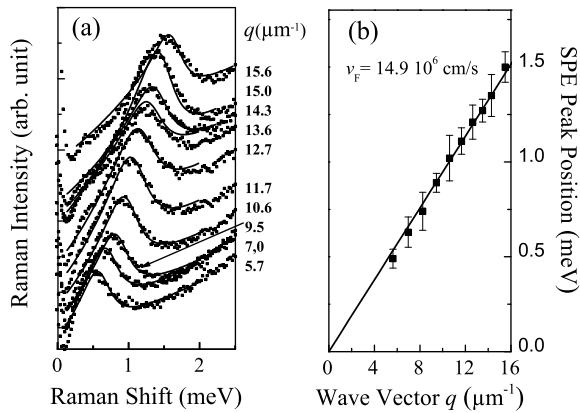}
\caption{\label{fig:density} (a) Raman scattering spectra (dots) by SPEs for different wave vectors $q$, and line 
fit using the Lindhard dielectric function (solid lines), for sample B. A background contribution is also included 
in the spectral line fits. (b) Peak positions of the SPE lines shown in (a), plotted as a function of $q$. The 
Fermi velocity is obtained from the slope in (b). The density and the Fermi energy $E_{\text{F, Raman}}(0)$ of the 
electron gas are then extracted from the line fitting in (a) and the Fermi velocity in (b) using the uncorrected mass 
(see text).}
\end{figure}

Fig.~\ref{fig:density}(a) shows Raman scattering spectra by SPEs in the polarized configuration, at various in-plane 
wave vectors for sample B. For a given wave vector $q$, the intensity of the SPE line extends from zero to a maximum 
with peak at $\hbar v_{\text{F}}q$, where $v_{\text{F}}$ is the Fermi velocity. The linear dependence of its frequency 
($= v_{\text{F}}q + \hbar q^2/2 m^*$) on the corresponding $q$, for small $q$ values, results in a slope which is 
proportional to $v_{\text{F}}$. We obtain $v_{\text{F}}$ in two ways. In the first, we take the value of the peak 
position of the SPE lines, shown in Fig.~\ref{fig:density}(a) for various wave vectors and plot them as a function of 
$q$ in Fig.~\ref{fig:density}(b). The slope of the resulting line gives $v_{\text{F}}$. Using the bare band-edge mass 
$m_b(= 0.105 m_o)$\cite{cynt3} obtained from cyclotron measurements, we determine the electron density $n_s$ and the 
Fermi energy, which we define as
\begin{equation}
E_{\text{F, Raman}}(0)=(1/2) m_b{v_{\text{F}}}^2 = {\hbar}^2 \pi n_s/m_b . \label{Fermi}  
\end{equation}

In the second approach for obtaining $v_{\text{F}}$, we use the Lindhard dielectric response function, 
including a broadening parameter ($\sim 0.12$ meV) to phenomenologically account for the finite lifetime of the 
electron states and numerically integrating over the Fermi disk as outlined in Refs.~\onlinecite{Bernard4} and
\onlinecite{Richards}. We again use $m_b$ in our fit. Note that the values deduced from the Lindhard fits [shown in 
Fig.~\ref{fig:density}(a) as solid lines] serve as checks for the values obtained from the SPE peak position. In Fig.~\ref{fig:density}(a), we describe the background upon which the SPE line rides, associated with the upper tail of the 
luminescence, as an exponential function, increasing towards lower absolute outgoing photon energies. This background 
was then added to the Lindhard function to enable a fit to our data, shown in the figure. From each fit, a value for 
$v_{\text{F}}$ was determined and an average was taken over the range of wave vectors explored. $E_{\text{F, 
Raman}}(0)$ and $n_s$ were then obtained using Eq.~(\ref{Fermi}). These values, from the Lindhard fit, are given in 
Table I for all samples studied. In the case of sample B, from the Lindhard fit, $n_s$ and $E_{\text{F, 
Raman}}(0)$ were found to be $2.9 \times 10^{11}$ cm$^{-2}$ and 6.5 meV, respectively. The values determined from the 
SPE peak position, for the same sample, for $n_s$ and $E_{\text{F, Raman}}(0)$ were $2.7\times 10^{11}$ cm$^{-2}$ and 
$6.2$ meV, respectively, such that there is a good agreement between the two methods. The peak position coincides with 
the maximum of the Lindhard dielectric response function and hence the good agreement with the line fitting is as 
expected for all samples, except sample E: $E_{\text{F, Raman}}(0)$ obtained from the Lindhard fit was $6.1$ 
meV while that obtained from the peak position was $5.5$ meV (see section V.A for further discussion and comparison 
with the Fermi energy value determined from the PL lineshape).  

\begin{figure*}
\includegraphics[width=0.8\linewidth]{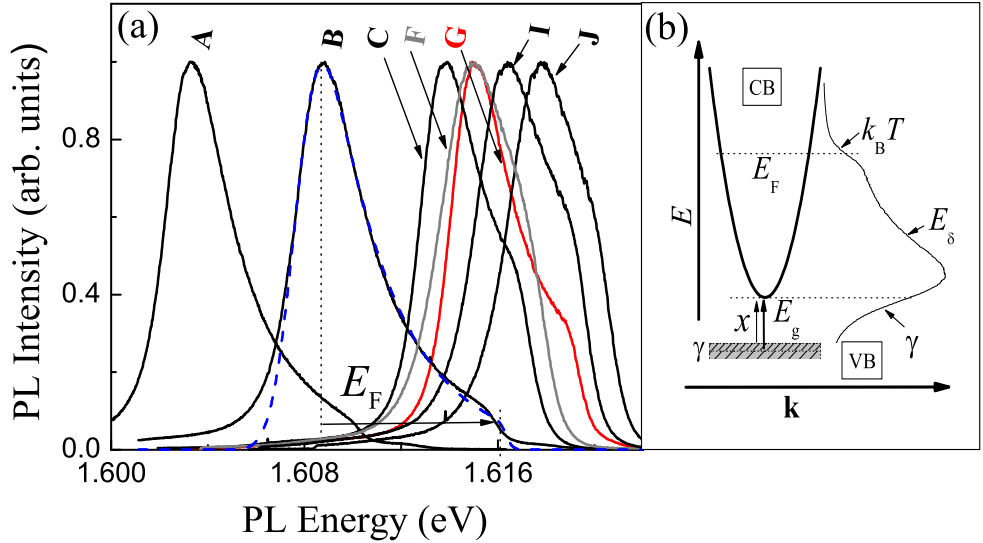}
\caption{\label{fig:pl} (a) (Color online) Normalized PL spectra for several chosen samples with different Mn$^{2+}$ 
concentration and electron densities. The Fermi energy values are extracted from the PL lineshape as explained in the 
text. The PL lineshape for sample B shows an approximate measure of the Fermi energy value (vertical dotted lines) 
given as $E_{\text{F, PL peak}}(0)$ in Table I and the PL lineshape fitting (blue dash line). (b) Schematics of the 
conduction band (CB) and valence band (VB) used in our fitting model given in Eq.~(\ref{PL}). A sample PL spectra is 
shown, depicting contributions to the various components in the expression (see text).}
\end{figure*}

\begin{table*}
\caption{\label{tab:table1} Sample Parameters. Values of $v_{\text{F, Raman}}$, $n_s$ and $E_{\text{F, Raman}}(0)$ were 
determined from the Lindhard lineshape fits, as described in section IV.A.1. Determination of $E_{\text{F, PL}}(0)$ 
is described in section IV.A.2. The manganese concentration is determined from the Raman measurements, described in 
IV.B.}
\begin{ruledtabular}
\begin{tabular}{lllllll}
Sample	& $n_s$ & $x$Mn & $v_{\text{F,Raman}}$ & $E_{\text{F, Raman}}(0)$  
     &	$E_{\text{F, PL peak}}(0)$ &$E_{\text{F, PL}}(0)$ \\
& ($\times 10^{11}\text{cm}^{-2}$) & (\%) & ($\times 10^{6}\text{cm/s}$) & (meV) & (meV) & (meV)\\
\hline
F	& 1.6 &	0.82 &	11.1&   3.6 &	3.0 & 4.2 \\
H	& 1.9 &	0.96 & 	12.1&   4.4 &	3.6 & 4.6 \\
C	& 2.1 &	0.79 &	12.7&   4.8 &	3.5 & 4.6 \\
I	& 2.2 &	0.96 &	13.0&   5.0 &	3.3 & 4.6 \\
D	& 2.3 &	0.78 &	13.3&   5.3 &	4.1 & 5.3 \\
G	& 2.3 &	0.84 &	13.4&   5.4 &	4.3 & 5.4 \\
J\footnotemark[1]	&  -  &	1.10  &	 -  &    -  &	2.8 & 4.2 \\ 
E	& 2.7 &	0.81 &	14.3&   6.1 &	3.8 & 5.2 \\	
B	& 2.9 & 0.75 &	14.9&   6.5 &	7.2 & 8.5 \\
A	& 3.0 &	0.46 &	15.1&   6.8 &	7.0 & 8.2 \\
\end{tabular}
\end{ruledtabular}
\footnotetext[1]{$v_{\text{F}}$ could not be obtained from Raman scattering.}
\end{table*}

\subsubsection{Fermi energy determination from the photoluminescence lineshape}

We now consider extracting $E_{\text{F, PL}}(0)$ from the PL lineshapes shown in Fig.~\ref{fig:pl}(a). The PL lineshape shows a characteristic peak or maximum, which without disorder is associated with the energy gap recombination. 
The intensity decreases gradually with increasing photon energy to form a shoulder and then an edge positioned at the 
energy gap plus the Fermi energy. By naively taking the difference between the PL maximum and the PL edge, as 
illustrated in Fig.~\ref {fig:pl}(a) for sample B, one obtains Fermi energy values that are less than those predicted 
by the Raman scattering measurements. These values are given in Table I and labeled $E_{\text{F, PL peak}}(0)$. For 
example, for sample C, we obtained $E_{\text{F, PL peak}}(0) = 3.5$ meV as compared to $E_{\text{F, Raman}}(0) = 4.8$ 
meV. The large error in estimating the Fermi energy is due to the precise determination of the fundamental band gap 
from the PL spectra. Broadening due to disorder causes the energy gap to be different (shift to lower energy) from the 
PL maximum; therefore, an adequate model is needed to extract $E_{\text{F, PL}}(0)$ from the PL lineshape.\cite
{ChristenTheory, Haufe, Perin} We perform such a PL lineshape fitting analysis using a phenomenological model 
described below. 

Christen and Bimberg\cite{ChristenTheory} reported a detailed calculation on PL profiles of quantum wells by 
considering, separately, free-electron, free-hole and excitonic recombination processes, and by accounting for 
broadening effects due to thermal distribution of carriers and interface roughness or composition fluctuations and 
final state recombination processes. In addition, the authors considered both wave vector conservation and 
non-conservation. In the same spirit, we present a phenomenological model that describes the PL lineshape of our 2DEG 
systems. We assume a parabolic band throughout this paper, neglect lifetime broadening due to final state 
recombination processes, and assume an infinite heavy-hole mass (so that the valence band is flat with respect to the 
conduction band), as the holes are localized. Note that for the electron densities considered in this paper, Coulomb 
interaction between the Fermi disk and the hole state, known to affect the PL lineshape for very low electron 
densities,\cite{Hourd} has been shown to be inhibited by screening and phase space filling; thus, the PL spectrum 
reproduces single-particle occupancy of the density of states.\cite{Hawrylak, Keller} We can, therefore, disregard 
excitonic effects in the PL lineshape. We treat separately potential and energy fluctuations due to disorder effects 
and assume that the Fermi energy is constant. For nonconservation of wave vector, an analogous form of equation 18(b) 
of Ref.~\onlinecite{ChristenTheory} for our PL lineshape is:
\begin{equation}
I(E)= A f(E-E_{\text{F}}-E_{\text{g}}) \int_0^\infty dx e^{(x-E_{\text{g}})^{2}\over(2\gamma^2)} \Theta(E-x) 
e^{(E-x)\over E_\delta}\label{PL}.
\end{equation}
Going from left to right of Eq.~(\ref{PL}), $A$ is a proportionality constant and $f(E-E_{\text{F}}-E_{\text{g}})$ is 
the Fermi distribution function, where $E_{\text{g}}$ is the average energy gap, $E_{\text{F}}$ is the Fermi energy and 
$E$ is the photon energy. The Gaussian envelope function with $\gamma$ accounts for broadening due to disorder effects 
on the low energy side of the PL spectrum, and it corresponds to the width of the fluctuations in energy of the hole 
band due to disorder. The hole state, because of its heavy mass, is readily sensitive to electrostatic potential 
fluctuations caused by disorder effects resulting in hole localization. Since the electron mass is smaller than the 
hole mass and screening effects dominate in the electron plane, we assume zero potential fluctuation for electrons.\cite
{britguy} The integral parameter $x$ is the local energy gap, the modified gap as a result of the fluctuating hole 
energy states. The unit step function $\Theta(\text{E}-x)$ is included to account for the electron and hole joint 
density of states. We have assumed an exponential decay function $e^{(E-x)\over E_\delta}$ for the wave vector break 
down since this function gives a better description of the high energy portion of the PL lineshape compared to a 
Lorentzian or Gaussian function.\cite{PericTh} $E_\delta$ is a constant associated with wave vector nonconservation 
and it relates to localization of the hole band in $k$ - space due to in-plane potential fluctuations.\cite{britguy} 
These contributions are depicted in Fig.~\ref{fig:pl}(b). 

The adjustable parameters used in fitting our data [shown in Fig.~\ref{fig:pl}(a)] in Eq.~(\ref{PL}) are $A$, 
$\gamma$, $E_{\text{g}}$, $E_\delta$ and $E_{\text{F}}$. We emphasize that the influence of these parameters on the 
lineshape fitting model is well separated so that each parameter is uniquely determined. We have used $T_h$ = $T_e$ = 
$T$ = 1.5 K, the experimental temperature. A sample fit is displayed in Fig.~\ref{fig:pl}(a) as blue dash line for 
sample B. The discrepancy between the fit and experiment at the Fermi edge results from the presence of a Fermi edge 
singularity. The values obtained for $E_{\text{F, PL}}(0)$ are given in Table I for all samples. The values for 
$E_\delta$, $E_g$ and $\gamma$ range from $\sim2$ to $\sim5$ meV, $1602$ to $1617$ meV and $\sim 0.7$ to $\sim 1.2$ 
meV, respectively. For such a simplified fitting model, we find that $E_{\text{F, PL}}(0)$ and $E_{\text{F, Raman}}(0)$
 are identical except for samples A, B, and E. For sample B, $E_{\text{F, PL}}(0) = 8.3$ meV, much larger than 
$E_{\text{F, Raman}}(0) = 6.5$ meV. We will discuss the results further in section IV.A. Next, we analyze the PL lineshape.

\subsubsection{Further analysis of the PL lineshape}

We comment on additional features of the PL lineshape. First note that, in Fig.~\ref{fig:pl}(a), the energy gap (near 
the PL maximum) increases in energy with increasing Mn$^{2+}$ concentration, going from sample A to sample J. The 
fitted values evolve from $1602$ to $1617$ meV. The dependence of the energy gap on Mn$^{2+}$ concentration has been 
reported by Matsuda $et$ $al.$\cite{Matsuda} on CdMnTe systems and independently by Kossut $et$ $al.$\cite{Kossut} 

Secondly, the PL lineshape gradually changes from sample A (large density) to sample J (low density), showing a 
gradual smearing of the shoulder near the Fermi-edge. Similar observations were made on 
$\text{Cd}_{1-x}\text{Mn}_x\text{Te}$ quantum wells, for example, in Ref.~\onlinecite{Jusserand1} and in Ref.~\onlinecite{Hourd}. The behavior was qualitatively attributed to potential fluctuations which causes non-vertical 
recombination between conduction electrons and photo-generated holes. Depending on the amplitude of the 
potential fluctuation, related to $E_\delta$ in our model, the PL line peaks at the zone center (as is the case for 
sample A) or extends to the PL edge (as is the case for sample J).\cite{Jusserand1, Hourd} This effect increases with 
decreasing electron density as observed from sample A to J due to increasing sensitivity of electrons to the potential 
fluctuation. We therefore fit the PL lineshape considering potential fluctuations in the electron plane. For sample 
A, we obtained $E_\delta \sim 2$ meV and for sample J, we obtained $E_\delta \sim 5$ meV. 

\subsection{Magnetic field effects}
We now present the magnetic field dependent results. As in the last section, we will start with Raman scattering 
measurements and later present the magneto-PL results. 

\subsubsection{Zeeman energies from Raman scattering}

\begin{figure}
\includegraphics[width=0.7\linewidth]{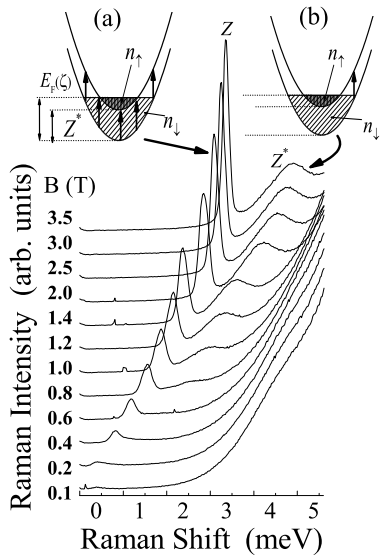}
\caption{\label{fig:MagnetoRam} Raman scattering measurements on sample B for various in-plane magnetic fields (Voigt 
configuration). Two lines corresponding to the collective and individual spin-flip excitations associated with the bare Zeeman splitting $Z$ and the renormalized Zeeman splitting $Z^*$, respectively, are shown, in accordance with Larmor$'$s theorem at zero wave vector. The inserts (a) and (b) show the spin-splitting of the conduction band with two spin populations: spin-up $n_{\uparrow}$ and spin-down $n_{\downarrow}$. Insert (a) shows the collective spin-flip of electrons from the spin-down subband to the spin-up subband, and insert (b) shows a single electron flipping its spin from spin-down to spin-up subband. The corresponding separation between the spin subbands yields $Z^*$.}
\end{figure}

Under a small magnetic field, the lowest conduction subband splits into two spin bands as shown in the inserts of Fig.~\ref{fig:MagnetoRam}. In the depolarized configuration, the SPE spectrum consists of excitations from the majority 
(electrons having spin-down with density $n_\downarrow$) to the minority spin subbands (electrons having spin-up with 
density $n_\uparrow$). The spin-flip Raman spectra in Fig.~\ref{fig:MagnetoRam} show two features below 6 meV: at $q = 
0$, these lines are attributed to the collective spin-flip excitation, insert (a), and the single-particle spin-flip 
excitation, insert (b), corresponding to the bare Zeeman splitting $Z$ and the renormalized Zeeman splitting $Z^*$, 
respectively.\cite{Jusserand4,Perez1} The energy value of $Z^*$ is greater than $Z$ because of the renormalization of 
the spin subband separation due to exchange-correlation interactions.\cite{Perez1, Florent2} With a knowledge of $Z^*$ and the Fermi energy obtained in section IV.A, the degree of spin-polarization in our quantum wells can be determined. Also, knowing $Z$, the Mn$^{2+}$ content of our quantum wells can be obtained. We shall discuss the spin polarization values deduced from Raman scattering in section V.C.  

\begin{figure}
\includegraphics[width=\linewidth]{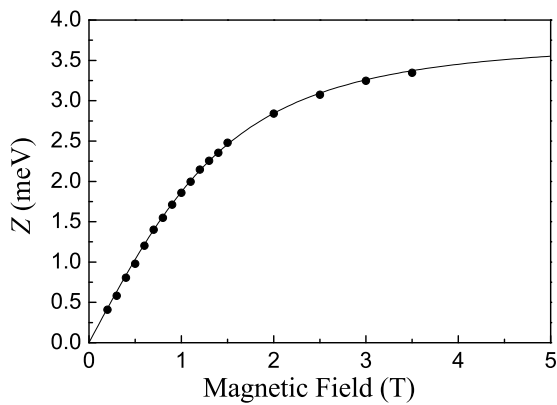}
\caption{\label{fig:MagnetoRam2} $Z$ values (spheres) obtained from Fig.~\ref{fig:MagnetoRam} and plotted as a 
function of the magnetic field for sample B. The black curve is a Brillouin function fitted to the data points. At a 
saturation magnetic field of 4 T, the Mn$^{2+}$ concentration extracted = 0.75\%, $Z = 3.5$ meV and the temperature 
$T_m$ = 1.5 K.}
\end{figure}

In Fig.~\ref{fig:MagnetoRam2}, we show the extraction of the Mn$^{2+}$ concentration for sample B using the modified
Brillouin function. The Brillouin function describes the thermodynamic average of the spin state of Mn$^{2+}$ 
ions and is related to $Z$ by $Z=-N_0 \alpha x \langle{S_z(B,T)}\rangle$, where $\alpha$ is the exchange integral, 
$N_0$ is the number of unit cells per unit volume, $\langle{S_z(B,T)}\rangle = S_o B_{5/2}({{5\over 2} g\mu_B B\over 
k_B(T+To)})$ is the Brillouin function, $x$ is the mole fraction of Mn$^{2+}$, the spin value is $\sim 5/2$ for our 
Mn$^{2+}$ concentration, $\mu_B$ is the Bohr magneton, $g = 2$, and $S_o$ and $T_o$ are parameters associated with the 
manganese atoms.\cite{gaj} The magnetic field dependence of $Z$ is well reproduced by the Brillouin curve (shown as a 
solid line) in the figure; this fit gives a Mn$^{2+}$ concentration of 0.75 \% and a temperature of 1.5 K for sample 
B. The manganese concentrations determined in this way are given in Table I for all samples. 

\subsubsection{Zeeman splitting from magneto-PL}

\begin{figure*}
\includegraphics[width=0.7\linewidth]{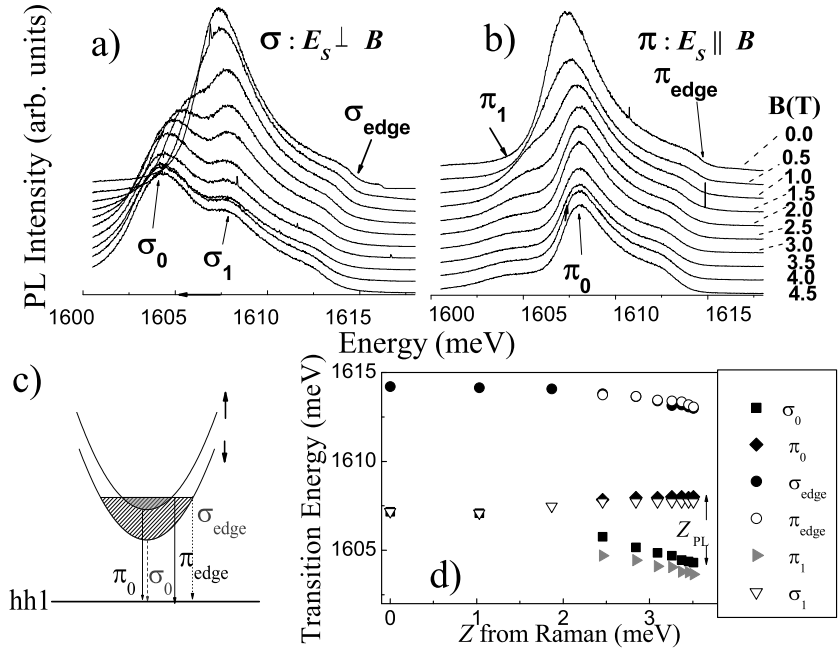}
\caption{\label{fig:magnetopl} PL spectra in the Voigt configuration on sample B at various magnetic fields and 1.5 K: 
(a) spectra from the majority spin-split subband taken with the in-plane magnetic field perpendicular to the 
polarization of the outgoing emission; (b) spectra from the minority spin-split subbands taken with the in-plane 
magnetic field parallel to the outgoing emission. (c) Schematic diagram showing the electronic transitions.  d) Points 
extracted from (a) and (b) for the following transitions: $\sigma_1$, $\sigma_0$, $\pi_0$ , $\pi_1$,  
$\sigma_{\text{edge}}$ and $\pi_{\text{edge}}$ depicted in (c). The $\pi_1$($\sigma_1$) transitions are identical to 
the $\sigma_0$ ($\pi_0$) transitions. The difference between the $\pi_0$ and the $\sigma_0$ transitions gives $Z_{\text{PL}}$, 
refer to Fig.~\ref{fig:MagnetoRam}}
\end{figure*}

\begin{figure}
\includegraphics[width=\linewidth]{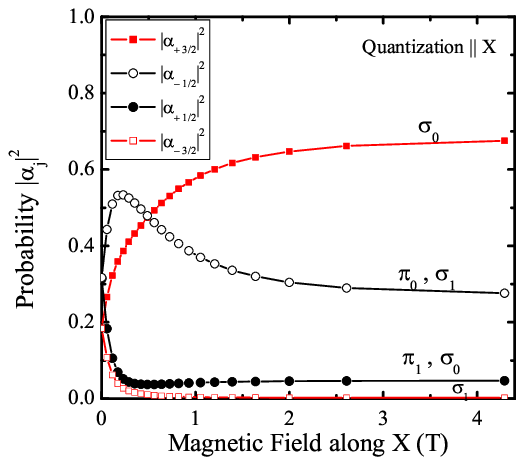}
\caption{\label{fig:PLCalc} (Color online) Decomposition of heavy-hole states in an applied magnetic field for sample B 
calculated for the Voigt configuration and a temperature $T$ = 1.5 K. Each curve represents the probability amplitude of heavy-hole states in our quantum wells.}
\end{figure}

In an applied magnetic field, electron and hole bands split; however, in the Voigt configuration, Zeeman splitting 
within the heavy-hole band is vanishingly small in small magnetic fields due to the hole spin alignment in the growth 
direction.\cite{Peyla} Fig.~\ref{fig:magnetopl} shows magneto-PL spectra on sample B in the Voigt configuration. 
Significant components of Fig.~\ref{fig:magnetopl}, at high magnetic field, are features associated with each spin 
population: Fig.~\ref{fig:magnetopl}(a), the majority spin-split subband for the spin-down electrons and, Fig.~\ref
{fig:magnetopl}(b), the minority spin-split subband for the spin-up electrons. We access these spin subbands by 
changing the polarization of the detected photons. To access the spin-down population, the outgoing laser beam 
polarization ($E_s$) is orthogonal to the applied magnetic field ($B$). This is the $\sigma$ polarization shown in 
Fig.~\ref{fig:magnetopl}(a) for various magnetic fields. For the spin-down population, the outgoing polarization is 
parallel to the applied magnetic field, shown in Fig.~\ref{fig:magnetopl}(b). The latter is the $\pi$ polarization. 
The schematic diagram shown in Fig.~\ref{fig:magnetopl}(c) is a depiction of the spin subbands and the electronic 
transitions corresponding to the spectra shown in Fig.~\ref{fig:magnetopl} (a) and (b). The edges of the PL are 
labeled $\sigma_{\text{edge}}$ and $\pi_{\text{edge}}$. 

For an understanding of Fig.~\ref{fig:magnetopl}, consider recombination processes occurring between the first 
heavy-hole band and the electrons. The valence band state for the heavy-hole can be expressed in terms of four 
component states:\cite{Peyla}
\begin{equation}
|hh_1\rangle  = \sum_J \alpha_J |J\rangle \label{hh},
\end{equation}
where $J = {\pm} {1\over 2}$; ${\pm} {3\over 2}$, and ${|\alpha_J|}^2$ are the probability amplitudes of the hole 
states relating to the transition amplitude and normalizing factors of the electron-hole recombination processes, calculated using the envelope function approximation and plotted in Fig.~\ref{fig:PLCalc}. In Fig.~\ref
{fig:PLCalc}, the quantization axis is chosen along the $x$-direction and the growth direction is in the $z$-direction. At high magnetic fields, the heavy-hole state aligns parallel to the field since the best quantization is 
along the field direction.\cite{Peyla} The heavy-hole wave function is then described by $|hh_1\rangle \simeq 
\alpha_{-{1\over 2}} 
|-{1\over 2}\rangle + \alpha_{3\over 2}|{3\over 2}\rangle$, where $\alpha_{-{1\over 2}}$ and $\alpha_{3\over 2}$ are 
shown in Fig.~\ref{fig:PLCalc}. Using well-known electric dipole transition selection rules for $\sigma$ and $\pi$ 
photons:\cite{Peyla}
\begin{equation}
\langle J'|P_{\pi}|J\rangle  = 2p \delta_{J,-J'},
\end{equation}
\begin{equation}
\langle J'|P_{\sigma}|J\rangle = p ( \delta_{J,-J'} + \surd {3} \delta_{J, -3J'}),
\end{equation}
where $p$ is a constant and $J' = {\pm} {1\over 2}$, we expect to see the following recombination 
processes (electrons to holes): $\pi_0$ = $|{1\over2}\rangle \Rightarrow 
|-{1\over2}\rangle$; $\sigma_0$ = $|-{1\over2}\rangle \Rightarrow |{1\over2}\rangle$, $\sigma_0$
 = $|-{1\over2}\rangle \Rightarrow |{3\over2}\rangle$ and $\sigma_1$ = $|{1\over2}\rangle \Rightarrow 
|-{1\over2}\rangle$
. The luminescence line is dominated by $\sigma_0$ and $\pi_0$ recombination processes at high and low magnetic 
fields. $\sigma_1$ has a lower contribution than $\sigma_0$ according to our calculations shown in Fig.~\ref
{fig:PLCalc}. For intermediate magnetic fields, contributions from $\alpha_{-{1\over 2}}$ and $\alpha_{3\over 2}$ 
increase and we expect to have additional processes $\pi_1$ = $|-{1\over2}\rangle \Rightarrow |{1\over2}\rangle$ 
occurring and a reinforcement of $\sigma_1$ = $|{1\over2}\rangle \Rightarrow |-{3\over2}\rangle$ occurring. The 
$\sigma_1$ and $\pi_1$ transitions are mirror images of the $\pi_0$ and $\sigma_0$ transitions, respectively, arising 
from the fact that the hole spin state is impure. These processes are in very good agreement with our data shown in 
Figs.~\ref{fig:magnetopl}(a) and Fig.~\ref{fig:magnetopl}(b).

The transition energies extracted from Fig.~\ref {fig:magnetopl}(a) and (b) are plotted in Fig.~\ref
{fig:magnetopl}(d) as a function of the bare Zeeman energy obtained from Raman scattering (from Fig.~\ref
{fig:MagnetoRam2}). Note that the PL edges in both $\sigma$ and $\pi$ polarizations and the values for $\sigma_1$ $(\pi_1)$ and $\pi_0 (\sigma_0)$ nearly overlap, showing good agreement. The separation between the spin-up and the 
spin-down subbands is equivalent to the Zeeman energy as depicted in Fig.~\ref{fig:magnetopl}(c). The difference between the transition $\sigma_0$ and $\pi_0$ is expected to equal $Z^*$ (for an interacting electron gas). 
In Fig.~\ref{fig:magnetopl}(d), however, the energy separation between the peak position $\sigma_0$ and the peak 
position $\pi_0$, is equivalent to $Z$ and not $Z^*$ contrary to our expectation. 

\begin{figure}
\includegraphics[width=\linewidth]{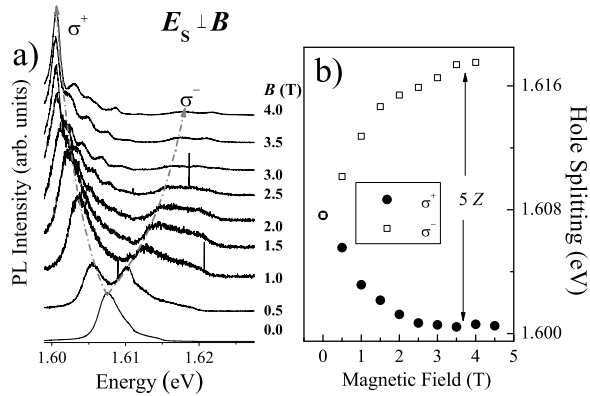}
\caption{\label{fig:farad} Magneto-PL measurements on sample B in the Faraday configuration. (a) Spectra taken at 
various magnetic fields at 1.5 K. (b) Transition energies extracted from (a).}
\end{figure}

To check the consistency of our observation for the Zeeman energy, measurements were taken in the Faraday geometry on 
sample B and this is shown in Fig.~\ref {fig:farad}(a). In the Faraday configuration, the magnetic field is parallel 
to the growth axes and photoluminescence from the heavy-hole and light hole-bands are known to split symmetrically 
into two circularly polarized components, $\sigma^+$ and $\sigma^-$ components.\cite{Peyla} In Fig.~\ref
{fig:farad}(a), these spectra, unlike the Voigt spectra, are mixed with Landau oscillations. The characteristic 
separation between the $\sigma^+$ and the $\sigma^-$ photoluminescence is equivalent to the heavy-hole splitting 
energy of 15.7 meV at 4 T, Fig.~\ref{fig:farad} (b). From experimentally determined values for the conduction band 
exchange integral ($N_0 \alpha = 0.22$ eV) and that for the valence band exchange integral ($N_0 \beta = -0.88 eV = 
4N_0 \alpha$), the heavy-hole splitting is equivalent to five times the conduction band splitting and therefore is 
expected to equal $5Z^*$.\cite{gaj} However, the Zeeman energy value obtained from Fig.~\ref{fig:farad}(b) is 
identical to $Z$, as was the case in the Voigt configuration. We will discuss the implication of this observation in 
section V.A and compare our measured values to the Raman scattering results.

\section{Discussion}
\subsection{The Fermi energy values}

\begin{figure}
\includegraphics[width=\linewidth]{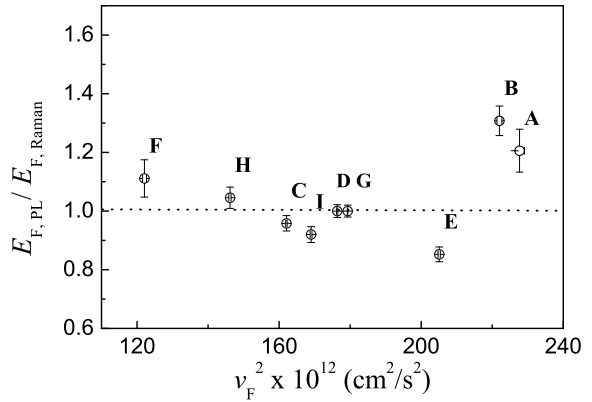}
\caption{\label{fig:ratioEF} Ratio of Fermi energy values obtained from Raman scattering and the PL lineshape 
fitting as a function of the Fermi velocity from Raman scattering. The vertical error bars are estimated from both PL 
and Raman scattering. The two results are identical for most samples except for samples A, B and E. The dash line 
in the figure shows identical results for both Raman scattering and PL.}
\end{figure}

To investigate the Fermi energy values obtained from the PL and Raman scattering measurements, we plot in Fig.~\ref
{fig:ratioEF} the ratio $E_{\text{F, PL}}(0)$ to $E_{\text{F, Raman}}(0)$ as a function of ${v_{\text{F}}}^2$ obtained 
from Raman scattering (refer to Table I). Recall that in section IV.A, the Fermi energy values are the same for most 
of our samples except for samples A, B, and E. The uncertainties in measuring the ratio of Fermi energies (by the 
Lindhard and the PL lineshape fitting analysis) are shown in Fig.~\ref{fig:ratioEF} as vertical error bars. The error 
is larger at low and large densities. Dispersive Raman scattering measurements are limited by disorder and for low 
electron densities ($< 1.5\times 10^{11}$ cm$^{-2}$), the SPE Raman line cannot be discernible since it gets too close 
to the excitation line and this will also affect the lineshape fitting. Although the PL lineshape fitting 
could not perfectly fit the high density samples near the Fermi edge, we can determine the Fermi energy with an 
error of less than 5\%. 

For samples A and B, the $E_{\text{F, Raman}}(0)$ values are smaller than those obtained from the PL lineshape while 
for sample E, the $E_{\text{F, Raman}}(0)$ value is larger. For sample E, due to large disorder evident in the PL lineshape (with $E_\delta \sim 4$ meV, and this is also the case for samples J), the SPE line was difficult to fit 
by the Lindhard model. The peak position values from Raman scattering however gave an $E_{\text{F, Raman}}(0)$ value 
of $5.5$ meV which agrees with $E_{\text{F, PL}}(0) = 5.2$ meV. For sample A and B, the estimated error in obtaining 
$E_{\text{F, PL}}(0)$ is less than the difference between the values for $E_{\text{F, Raman}}(0)$ and $E_{\text{F, 
PL}}(0)$, we note that the measured densities from both Raman scattering and PL are large.   

PL provides information on the entire curvature of the electron band. Hence, information on the renormalized mass 
should be accounted for in the PL lineshape. This means that $E_{\text{F, PL}}(0)$ includes the mass correction and 
should provide an accurate estimation of the Fermi energy within the parabolic band approximation (and assuming 
negligible contributions from Coulomb interaction between electrons and hole).\cite{Hawrylak} The Raman scattering 
process probes a limited region of the electronic band close to the Fermi energy, meaning that Raman scattering 
measurements are accurate in the determination of $v_{\text{F}}$. 

In Fig.~\ref{fig:ratioEF}, the ratio $E_{\text{F, PL}}(0) / {E_{\text{F, Raman}}(0)}$, assuming Eq.~(\ref{Fermi}) for 
$E_{\text{F, Raman}}(0)$ and $E_{\text{F, PL}}(0)= (1/2)m^* {v_{\text{F}}}^2$, is equal to the ratio $m^*/m_b$. In 
$\text{Cd}_{1-x}\text{Mn}_x\text{Te}$ quantum wells, the available information on the electron mass is that measured 
for CdTe by cyclotron resonance experiments, $m_b$. This mass does not include additional corrections due to many-body 
electron-electron interaction and gives only the bare mass.\cite{Kohn} As we have assumed a value for this bare mass
($m_b = 0.105 m_0$), the absolute values found for $m^*/m_b$ are unclear, but the qualitative behavior of $m^*/m_b$ 
with a minimum around $v_{\text{F}}= 13.4$ ($r_s = 2.3$) resembles calculations of Fig.10 in Ref.~\onlinecite{Asgari}. 
Nonetheless, we estimate less than 20\% increase in mass.

Our results also reaffirm that extracting the Fermi energy from the PL maximum and the Fermi edge of quantum wells 
gives inaccurate results and that a good lineshape fitting analysis is essential. Further, if we correct the 
values for the densities in Table I, deduced from $v_{\text{F}}$ assuming a bare mass, with the mass renormalization 
found in Fig.~\ref{fig:ratioEF}, we find for sample B $n_s = 4.9$x$ 10^{11}$ cm$^{-2}$ instead of $2.9$x$10^{11}$ cm
$^{-2}$.

\subsection{Collective and single-particle behavior}

\begin{figure}
\includegraphics[width=\linewidth]{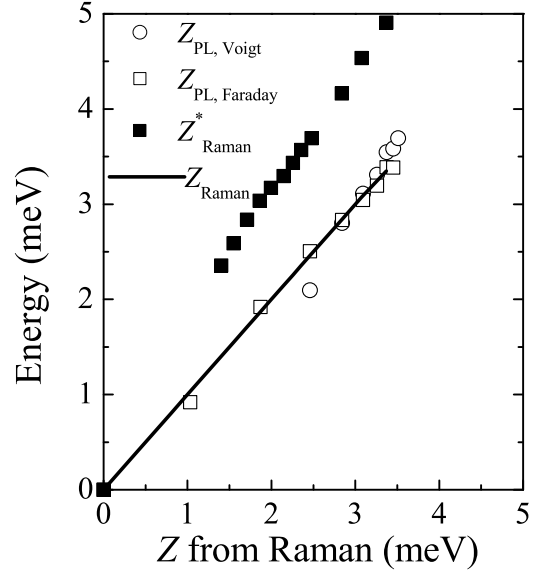}
\caption{\label{fig:Z} Energy values for $Z^*$ (solid squares) from the Raman scattering spectra shown in Fig.~\ref
{fig:MagnetoRam} and $Z_{\text{PL}}$ from the PL lineshape in the Voigt (open circles) from Fig.~\ref
{fig:magnetopl} and in the Faraday configurations (open squares) from Fig.~\ref{fig:farad} for sample B plotted 
against $Z$ from Raman scattering (see Fig.~\ref{fig:MagnetoRam}). The $Z_{\text{PL}}$ values lie on the solid line 
indicating that they are equivalent to $Z$ from Raman scattering.}
\end{figure}

We have shown in section IV.B.1 that the Raman scattering measurements identify both collective and single-particle 
behavior of a spin-polarized 2DEG, whereas for the magneto-PL measurements given in section IV.B.2, only the bare 
Zeeman splitting $Z$ was extracted. Values of $Z$ obtained from the PL lineshape in the Voigt and Faraday 
configurations, as well as $Z^*$ from Raman scattering are plotted in Fig.~\ref{fig:Z} as a function of $Z$ from Raman 
scattering. In that figure, the $Z$ values extracted from the PL measurements are consistent with the Raman scattering 
values. The analysis was repeated for the samples listed in Table I and the results were found to be consistent: that 
is, the PL lineshape gives only $Z$ and not $Z^*$. As $Z$ is associated with a spin-flip energy related to collective 
excitations, we might conclude that collective effects influence the PL behavior.

\begin{figure}
\includegraphics[width=\linewidth]{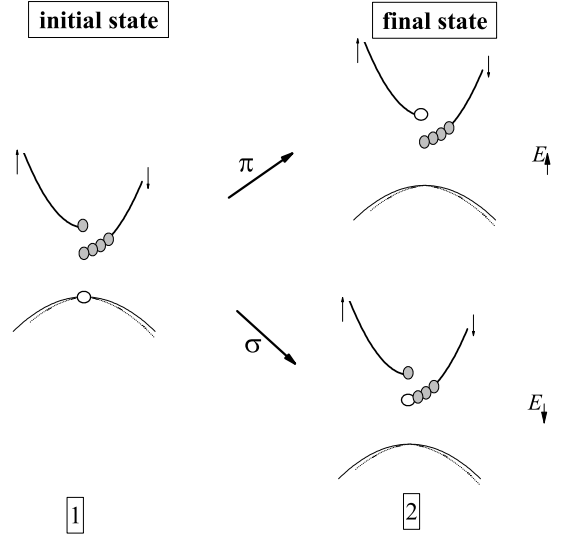}
\caption{\label{fig:colfig} Schematic diagram of the optical transition processes occurring in Fig.~\ref
{fig:magnetopl}. Panel (1) is the initial state and panel (2) is the final state. Electrons are represented by solid spheres and holes are represented by open circles.}
\end{figure}
To understand the magneto-PL behavior, let us consider the various transition processes that occur. We follow 
the evolution of the PL profile, shown in Fig.~\ref{fig:colfig} following the work of Kossacki and co-authors for holes.
\cite{Kossacki} This figure shows the representation of many-body states in a quantum well. In the initial state, 
shown in panel (1), a photon is absorbed by an electron creating a hole in the first heavy-hole spin split 
subband. In the final state, panel (2), a photon is emitted with two possibilities arising: a) an electron in 
the spin-up band recombines with the hole or b) an electron in the spin-down band recombines with the hole leaving an 
unfilled space in either spin-up or spin-down bands. The difference between the energies of the spin-up ($E_\uparrow$) 
and spin-down ($E_\downarrow$) transitions, which are shown experimentally in Fig.~\ref{fig:magnetopl}(b) for the 
$\pi_0$ transition and Fig.~\ref{fig:magnetopl}(a) for the $\sigma_0$ transition of section IV.B.2, is exactly the 
bare Zeeman splitting energy. This is possible, if, the final state in the transition $E_\downarrow$ is an 
excited state of a $q = 0$ spin-flip wave whose energy is $Z$. Hence, the difference between the transitions for the 
spin-up and the spin-down subbands yield $Z$ and not $Z^*$. We therefore conclude that the PL spectra is 
insensitive to the exchange modified spin splitting $Z^*$ but exhibit collective recombination processes. 
Additionally, we conclude that Raman scattering provides a better measure of both the Zeeman splitting $Z$ and the 
modified Zeeman energy due to Coulomb interactions $Z^*$.\cite{Perez1, Jusserand4, Florent2} 

\subsection{Determination of the spin polarization degree from Raman scattering and PL}

We now turn to the determination of the spin polarization degree $\zeta$ from Raman scattering and PL. Consider the 
magneto-PL measurements in Fig.~\ref{fig:magnetopl} of section IV.B.2. Assuming a two-dimensional parabolic band for 
each spin-split subband and the same mass renormalization for each, then from the PL lineshape, the density of 
spin-up electrons $n_{\uparrow}\propto E_{\uparrow}$ and that for spin-down electrons $n_{\downarrow}\propto 
E_{\downarrow}$ where $E_{\downarrow} = \sigma_{edge} - \sigma_0 - \delta w$ and $E_{\uparrow} = \pi_{edge} - 
\pi_0-\delta w$ are the PL linewidths in the $\sigma$ and the $\pi$ polarizations, respectively. Here $\delta w$ (= PL 
peak - $E_g$) is broadening due to disorder effects on the low energy side of the magneto-PL spectra (see Fig.~\ref
{fig:magnetopl}(a) and (b)). For an in-plane magnetic field, the hole orbit is not quantized, implying that the 
localization length is unchanged. This assumes that broadening due to disorder, localized in the hole band, is 
independent of the applied magnetic field while localization occurs in plane.\cite{britguy} Since the heavy-hole 
splitting is small in the Voigt configuration, we can assume the same disorder effect for both heavy-hole spin split 
bands. We therefore approximate the width contribution $\delta w$ to $\gamma$ (the energy fluctuation parameter for 
zero magnetic field disorder contribution to the PL width obtained in section IV.A). 

\begin{figure}
\includegraphics[width=\linewidth]{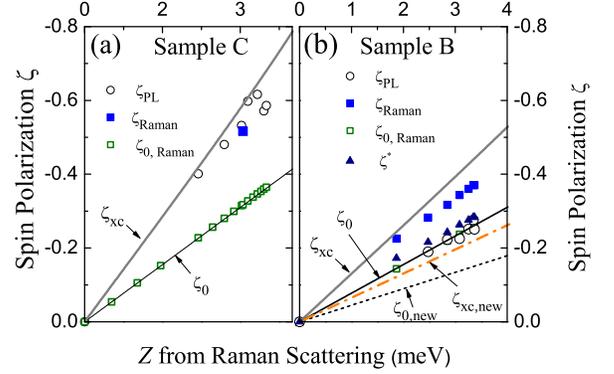}
\caption{\label{fig:zeta} (color online) $\zeta$ values obtained from PL and Raman scattering for (a) sample C and 
(b) sample B. $\zeta_{0,Raman}=Z_{Raman}/2 E_{\text{F, Raman}}$, $\zeta_{\text{Raman}}= (Z^*)_{Raman}/2 E_{\text{F, 
Raman}}$ and $\zeta_{PL}$ was obtained from the width of the PL in the $\sigma$ and the $\pi$ configuration (see Fig.~\ref{fig:magnetopl}). $\zeta_{xc}$ and $\zeta_{0}$ (solid black lines) are the theoretical values for the interacting  
and bare spin polarization degrees, respectively. $\zeta_{xc, new}$ and $\zeta_{0, new}$ (orange dot-dash curve and 
black dotted line) are the corrected values obtained using the mass correction of Fig.~\ref{fig:ratioEF}.}
\end{figure}

From Eq.~(\ref{Zeta}) in the introduction, $\zeta$ can be rewritten as  
\begin{equation}
(E_{\uparrow}-E_{\downarrow})/(E_{\uparrow}+E_{\downarrow})\label{Zeta2}.
\end{equation}
The values of $\zeta$ obtained by this means, labeled $\zeta_{\text{PL}}$, assume an equal mass renormalization for 
both spin population, independent of the spin polarization degree. This is certainly valid for intermediate spin 
polarization degree.\cite{DasSarma2} These values are plotted in Fig.~\ref{fig:zeta}(a) for sample C and in Fig.~\ref
{fig:zeta}(b) for sample B as a function of the bare Zeeman energy $Z$ from Raman scattering. We also define a spin 
polarization degree based on Raman scattering values as $\zeta_{\text{Raman}} = - Z^*_{\text{Raman}} / 2 E_{\text{F, 
Raman}}(0)$ and spin polarization degree based on both Raman scattering and PL defined as $\zeta^* = Z^*/2 E_{\text{F, 
PL}}(0)$, where we have used the Fermi energy value obtained from the PL lineshape fitting and $Z^*$ from Raman scattering. Here $E_{\text{F, Raman}}(0)$ [and similarly $E_{\text{F, PL}}(0)$] are the zero magnetic field 
values obtained in section IV.A. $\zeta^*$ is shown as blue solid triangles in Fig.~\ref{fig:zeta}(b) and is found to 
be comparable to $\zeta_{\text{PL}}$ in both samples.

To gain insight, we make a comparison with an exact theory of the spin polarization degree.\cite{Florent2, Perez1} For 
this, we define the bare spin polarization $\zeta_0 = - {Z m_b}/{\hbar^2 (2\pi n_s)}$ of a non-interacting electron 
gas and its corresponding value $\zeta_{xc}$ for the interacting case. The former quantity is expected to match 
$\zeta_{\text{0,Raman}} = Z_{\text{Raman}}/2 E_{\text{F, Raman}}(0)$ as $E_{\text{F, Raman}}(0)$ is defined using the 
bare mass. The latter quantity $\zeta_{xc}$ is the real spin polarization degree of the spin polarized 2DEG and is 
known to be enhanced due to exchange and correlation effects, and corrected for finite thickness of the quantum well.
\cite{Attaccalite, DePalo} We compare in Fig.\ref{fig:zeta} the theoretical values for the spin polarization degrees 
$\zeta_{xc}$ and $\zeta_0$ with the experimental ones as a function of $Z$. To calculate the theoretical ratio 
$\zeta_{xc}/\zeta_0$, we have used the model of Ref.~\onlinecite{Florent2} and the densities given in Table I.

In Fig.\ref{fig:zeta}(a) for sample C, the PL values (as well as the Raman values) for the spin polarization degree 
agree well with the model. The good agreement validates the fact that the Fermi energy values obtained from Raman 
scattering and PL are identical for sample C. In the case of sample B in Fig.~\ref{fig:zeta}(b), the       
$\zeta_{\text{PL}}$ values are lower than the Raman scattering values and all values (including the 
$\zeta_{\text{Raman}}$ values) are in poor agreement with the model. We suspect the strong mass renormalization 
found in Fig.~\ref{fig:ratioEF} for this sample (and sample A) to be at the origin of this inconsistency. Indeed, 
densities given in Table I are determined from $v_{\text{F}}$, which does not take into account the mass 
renormalization found in Fig.~\ref{fig:ratioEF}. If we now correct all densities for the mass, hence the new density 
$n_{2D} =(m^*/m_b)^2*n_s$, and recalculate the spin polarization degrees with the new densities, we find no change for 
sample C, but significant changes for sample B (shown as orange dot-dash line labeled $\zeta_{xc, new}$ and black 
dotted line labeled $\zeta_{0, new}$ in Fig.\ref{fig:zeta}), such that the results become consistent. Hence, the 
calculated renormalized spin polarization degree is now comparable with $\zeta_{PL}$ and $\zeta^*$.

We conclude that that $\zeta_{\text{PL}}$ and $\zeta^*$ are reliable estimations of the spin-polarization degree. 
Furthermore, our results suggest that the Fermi energy values obtained from the PL lineshape fitting analysis can be 
trusted. On the other hand, the Raman scattering determination of the Fermi energy and $\zeta$ using the bare mass, 
has to be corrected for the mass renormalization. 
 
Finally, we point out that in Fig.~\ref{fig:zeta} and for the same manganese concentration, the spin polarization 
degree shows an inverse relationship with the electron density, when comparing Fig.~\ref{fig:zeta}(a) and Fig.~\ref
{fig:zeta}(b) on the same vertical scale. This behavior is predicted rather well by the Raman scattering and PL 
results.  

\section{Conclusions}
We have performed a comparative study on a spin-polarized 2DEG using resonant electronic Raman scattering and PL. 
Three key observations were made after measuring the Fermi velocity, the Fermi energy, the Zeeman energies and the 
spin-polarization degree. First, assuming a parabolic band, the Fermi velocity and the Fermi energy measured from 
Raman scattering was comparable to that measured from the PL lineshape for moderate electron densities, whereas for 
large electron densities, the values differ unless a renormalized mass was introduced. Secondly, discrepancy in the 
spin polarization degree $\zeta$ also occurs for the samples showing this difference in the Fermi energy value. The 
discrepancies are attributed to the mass renormalization, which we have defined as the ratio between the Fermi 
energies extracted from PL and that deduced from Raman scattering. We checked the consistency of this result by a 
comparison with an exact model of the spin polarization. A good agreement was obtained after the density, extracted 
from $v_{\text{F}}$, was corrected for the mass renormalization. Our results suggest that the effective mass is 
modified in the presence of many-body electron-electron interaction for the range of $r_s$ values studied, though this 
range is small. 

Thirdly, the magneto-PL lineshape gives $Z$ instead of $Z^*$, contradictory to what we expect. Using a 
phenomenological analysis, we showed that the PL lineshape is insensitive to the exchange modified spin splitting $Z^*$ but is influenced by collective effects. 

From the above analysis, we conclude that Raman scattering determines $v_{\text{F}}$ and directly measures the bare 
and the renormalized Zeeman splitting accurately while PL provides accurate determination of the Fermi energy and the 
bare Zeeman splitting energy. In addition, the PL lineshape gives a good estimate of the renormalized spin polarization for low spin polarized systems, while for the high spin polarized case, determination of the spin polarization degree requires knowledge of $Z^*$ from Raman scattering and the Fermi energy from PL.

\begin{acknowledgments}
The authors would like to thank M. Perrin for useful discussion. CA would like to thank M. Teichmann for technical 
help. This work was supported by the EPSRC and the CNRS. 
\end{acknowledgments}


\begin{thebibliography}{51}
\expandafter\ifx\csname natexlab\endcsname\relax\def\natexlab#1{#1}\fi
\expandafter\ifx\csname bibnamefont\endcsname\relax
  \def\bibnamefont#1{#1}\fi
\expandafter\ifx\csname bibfnamefont\endcsname\relax
  \def\bibfnamefont#1{#1}\fi
\expandafter\ifx\csname citenamefont\endcsname\relax
  \def\citenamefont#1{#1}\fi
\expandafter\ifx\csname url\endcsname\relax
  \def\url#1{\texttt{#1}}\fi
\expandafter\ifx\csname urlprefix\endcsname\relax\def\urlprefix{URL }\fi
\providecommand{\bibinfo}[2]{#2}
\providecommand{\eprint}[2][]{\url{#2}}

\bibitem[{\citenamefont{Karczewski et~al.}(1998)\citenamefont{Karczewski,
  Jaroszynski, Barcz, Kutrowski, Wojtowicz, and Kossut}}]{karczewski}
\bibinfo{author}{\bibfnamefont{G.}~\bibnamefont{Karczewski}},
  \bibinfo{author}{\bibfnamefont{J.}~\bibnamefont{Jaroszynski}},
  \bibinfo{author}{\bibfnamefont{A.}~\bibnamefont{Barcz}},
  \bibinfo{author}{\bibfnamefont{M.}~\bibnamefont{Kutrowski}},
  \bibinfo{author}{\bibfnamefont{T.}~\bibnamefont{Wojtowicz}},
  \bibnamefont{and} \bibinfo{author}{\bibfnamefont{J.}~\bibnamefont{Kossut}},
  \bibinfo{journal}{J. Cryst. Growth} \textbf{\bibinfo{volume}{184/185}},
  \bibinfo{pages}{814} (\bibinfo{year}{1998}).

\bibitem[{\citenamefont{Wolf et~al.}(2001)\citenamefont{Wolf, Awschalom,
  Buhrman, Daughton, von Molnar, Roukes, Chtchelkanova, and
  Treger}}]{Awschalom1}
\bibinfo{author}{\bibfnamefont{S.~A.} \bibnamefont{Wolf}},
  \bibinfo{author}{\bibfnamefont{D.~D.} \bibnamefont{Awschalom}},
  \bibinfo{author}{\bibfnamefont{R.~A.} \bibnamefont{Buhrman}},
  \bibinfo{author}{\bibfnamefont{J.~M.} \bibnamefont{Daughton}},
  \bibinfo{author}{\bibfnamefont{S.}~\bibnamefont{von Molnar}},
  \bibinfo{author}{\bibfnamefont{M.~L.} \bibnamefont{Roukes}},
  \bibinfo{author}{\bibfnamefont{A.~Y.} \bibnamefont{Chtchelkanova}},
  \bibnamefont{and} \bibinfo{author}{\bibfnamefont{D.~M.}
  \bibnamefont{Treger}}, \bibinfo{journal}{Science}
  \textbf{\bibinfo{volume}{294}}, \bibinfo{pages}{1488} (\bibinfo{year}{2001}).

\bibitem[{\citenamefont{Gaj et~al.}(1979)\citenamefont{Gaj, Planel, and
  Fishman}}]{gaj}
\bibinfo{author}{\bibfnamefont{J.}~\bibnamefont{Gaj}},
  \bibinfo{author}{\bibfnamefont{R.}~\bibnamefont{Planel}}, \bibnamefont{and}
  \bibinfo{author}{\bibfnamefont{G.}~\bibnamefont{Fishman}},
  \bibinfo{journal}{Solid State Commun.} \textbf{\bibinfo{volume}{29}},
  \bibinfo{pages}{435} (\bibinfo{year}{1979}).

\bibitem[{\citenamefont{Pinczuk et~al.}(1989)\citenamefont{Pinczuk,
  Schmitt-Rink, Danan, Valladares, Pfeiffer, and West}}]{Pinczuk}
\bibinfo{author}{\bibfnamefont{A.}~\bibnamefont{Pinczuk}},
  \bibinfo{author}{\bibfnamefont{S.}~\bibnamefont{Schmitt-Rink}},
  \bibinfo{author}{\bibfnamefont{G.}~\bibnamefont{Danan}},
  \bibinfo{author}{\bibfnamefont{J.~P.} \bibnamefont{Valladares}},
  \bibinfo{author}{\bibfnamefont{L.~N.} \bibnamefont{Pfeiffer}},
  \bibnamefont{and} \bibinfo{author}{\bibfnamefont{K.~W.} \bibnamefont{West}},
  \bibinfo{journal}{Phys. Rev. Lett.} \textbf{\bibinfo{volume}{63}},
  \bibinfo{pages}{1633} (\bibinfo{year}{1989}).

\bibitem[{\citenamefont{Pinczuk et~al.}(1971)\citenamefont{Pinczuk, Brillson,
  Burstein, and Anastassakis}}]{Pinczuk2}
\bibinfo{author}{\bibfnamefont{A.}~\bibnamefont{Pinczuk}},
  \bibinfo{author}{\bibfnamefont{L.}~\bibnamefont{Brillson}},
  \bibinfo{author}{\bibfnamefont{E.}~\bibnamefont{Burstein}}, \bibnamefont{and}
  \bibinfo{author}{\bibfnamefont{E.}~\bibnamefont{Anastassakis}},
  \bibinfo{journal}{Phys. Rev. Lett.} \textbf{\bibinfo{volume}{27}},
  \bibinfo{pages}{317} (\bibinfo{year}{1971}).

\bibitem[{\citenamefont{Tan et~al.}(2005)\citenamefont{Tan, Zhu, Stormer,
  Pfeiffer, Baldwin, and West}}]{Tan1}
\bibinfo{author}{\bibfnamefont{Y.-W.} \bibnamefont{Tan}},
  \bibinfo{author}{\bibfnamefont{J.}~\bibnamefont{Zhu}},
  \bibinfo{author}{\bibfnamefont{H.~L.}~\bibnamefont{Stormer}},
  \bibinfo{author}{\bibfnamefont{L.~N.} \bibnamefont{Pfeiffer}},
  \bibinfo{author}{\bibfnamefont{K.~W.} \bibnamefont{Baldwin}},
  \bibnamefont{and} \bibinfo{author}{\bibfnamefont{K.~W.} \bibnamefont{West}},
  \bibinfo{journal}{Phys. Rev. Lett.} \textbf{\bibinfo{volume}{94}},
  \bibinfo{pages}{016405} (\bibinfo{year}{2005}), \bibinfo{note}{and references
  therein}.

\bibitem[{\citenamefont{Smith et~al.}(1992)\citenamefont{Smith, MacDonald, and
  Gumbs}}]{Smith}
\bibinfo{author}{\bibfnamefont{A.~P.} \bibnamefont{Smith}},
  \bibinfo{author}{\bibfnamefont{A.~H.} \bibnamefont{MacDonald}},
  \bibnamefont{and} \bibinfo{author}{\bibfnamefont{G.}~\bibnamefont{Gumbs}},
  \bibinfo{journal}{Phys. Rev. B} \textbf{\bibinfo{volume}{45}},
  \bibinfo{pages}{8829} (\bibinfo{year}{1992}).

\bibitem[{\citenamefont{Asgari and Tanatar}(2006)}]{Asgari}
\bibinfo{author}{\bibfnamefont{R.}~\bibnamefont{Asgari}} \bibnamefont{and}
  \bibinfo{author}{\bibfnamefont{B.}~\bibnamefont{Tanatar}},
  \bibinfo{journal}{Phys. Rev. B} \textbf{\bibinfo{volume}{74}},
  \bibinfo{pages}{075301} (\bibinfo{year}{2006}).

\bibitem[{\citenamefont{Jusserand et~al.}(2003)\citenamefont{Jusserand, Perez,
  Richards, Karczewski, Wojtowicz, Testelin, Wolverson, and
  Davies}}]{Jusserand4}
\bibinfo{author}{\bibfnamefont{B.}~\bibnamefont{Jusserand}},
  \bibinfo{author}{\bibfnamefont{F.}~\bibnamefont{Perez}},
  \bibinfo{author}{\bibfnamefont{D.~R.} \bibnamefont{Richards}},
  \bibinfo{author}{\bibfnamefont{G.}~\bibnamefont{Karczewski}},
  \bibinfo{author}{\bibfnamefont{T.}~\bibnamefont{Wojtowicz}},
  \bibinfo{author}{\bibfnamefont{C.}~\bibnamefont{Testelin}},
  \bibinfo{author}{\bibfnamefont{D.}~\bibnamefont{Wolverson}},
  \bibnamefont{and} \bibinfo{author}{\bibfnamefont{J.~J.}
  \bibnamefont{Davies}}, \bibinfo{journal}{Phys. Rev. Lett.}
  \textbf{\bibinfo{volume}{91}}, \bibinfo{pages}{086802}
  (\bibinfo{year}{2003}).

\bibitem[{\citenamefont{Peric et~al.}(1993)\citenamefont{Peric, Jusserand,
  Richards, and Etienne}}]{Peric}
\bibinfo{author}{\bibfnamefont{H.}~\bibnamefont{Peric}},
  \bibinfo{author}{\bibfnamefont{B.}~\bibnamefont{Jusserand}},
  \bibinfo{author}{\bibfnamefont{D.}~\bibnamefont{Richards}}, \bibnamefont{and}
  \bibinfo{author}{\bibfnamefont{B.}~\bibnamefont{Etienne}},
  \bibinfo{journal}{Phys. Rev. B} \textbf{\bibinfo{volume}{47}},
  \bibinfo{pages}{12722} (\bibinfo{year}{1993}).

\bibitem[{\citenamefont{Kossacki et~al.}(2004)\citenamefont{Kossacki, Boukari,
  Bertolini, Ferrand, Cibert, Tatarenko, Gaj, Deveaud, Ciulin, and
  Potemski}}]{Kossacki}
\bibinfo{author}{\bibfnamefont{P.}~\bibnamefont{Kossacki}},
  \bibinfo{author}{\bibfnamefont{H.}~\bibnamefont{Boukari}},
  \bibinfo{author}{\bibfnamefont{M.}~\bibnamefont{Bertolini}},
  \bibinfo{author}{\bibfnamefont{D.}~\bibnamefont{Ferrand}},
  \bibinfo{author}{\bibfnamefont{J.}~\bibnamefont{Cibert}},
  \bibinfo{author}{\bibfnamefont{S.}~\bibnamefont{Tatarenko}},
  \bibinfo{author}{\bibfnamefont{J.~A.} \bibnamefont{Gaj}},
  \bibinfo{author}{\bibfnamefont{B.}~\bibnamefont{Deveaud}},
  \bibinfo{author}{\bibfnamefont{V.}~\bibnamefont{Ciulin}}, \bibnamefont{and}
  \bibinfo{author}{\bibfnamefont{M.}~\bibnamefont{Potemski}},
  \bibinfo{journal}{Phys. Rev. B} \textbf{\bibinfo{volume}{70}},
  \bibinfo{pages}{195337} (\bibinfo{year}{2004}).

\bibitem[{\citenamefont{Masalana et~al.}(2006)\citenamefont{Masalana, Kossacki,
  Pochocka, Golnik, Gaj, Ferrand, Bertolini, Tatarenko, and Cibert}}]{Masalana}
\bibinfo{author}{\bibfnamefont{W.}~\bibnamefont{Masalana}},
  \bibinfo{author}{\bibfnamefont{P.}~\bibnamefont{Kossacki}},
  \bibinfo{author}{\bibfnamefont{P.}~\bibnamefont{Pochocka}},
  \bibinfo{author}{\bibfnamefont{A.}~\bibnamefont{Golnik}},
  \bibinfo{author}{\bibfnamefont{J.~A.} \bibnamefont{Gaj}},
  \bibinfo{author}{\bibfnamefont{D.}~\bibnamefont{Ferrand}},
  \bibinfo{author}{\bibfnamefont{M.}~\bibnamefont{Bertolini}},
  \bibinfo{author}{\bibfnamefont{S.}~\bibnamefont{Tatarenko}},
  \bibnamefont{and} \bibinfo{author}{\bibfnamefont{J.}~\bibnamefont{Cibert}},
  \bibinfo{journal}{Appl. Phys. Lett.} \textbf{\bibinfo{volume}{89}},
  \bibinfo{pages}{052104} (\bibinfo{year}{2006}).

\bibitem[{\citenamefont{Jusserand et~al.}(2007)\citenamefont{Jusserand, Perrin,
  Lema\^itre, and Bloch}}]{Bernard4}
\bibinfo{author}{\bibfnamefont{B.}~\bibnamefont{Jusserand}},
  \bibinfo{author}{\bibfnamefont{M.}~\bibnamefont{Perrin}},
  \bibinfo{author}{\bibfnamefont{A.}~\bibnamefont{Lema\^itre}},
  \bibnamefont{and} \bibinfo{author}{\bibfnamefont{J.}~\bibnamefont{Bloch}}, in
  \emph{\bibinfo{booktitle}{Proceedings of the International Conference on the
  Physics of Semiconductors 28}}, edited by
  \bibinfo{editor}{\bibfnamefont{W.}~\bibnamefont{Jantsch}} \bibnamefont{and}
  \bibinfo{editor}{\bibfnamefont{F.}~\bibnamefont{Sch$\ddot{a}$ffler}}
  (\bibinfo{publisher}{AIP Conf. Proc.}, \bibinfo{year}{2007}), vol.
  \bibinfo{volume}{893}, p. \bibinfo{pages}{411}.

\bibitem[{\citenamefont{Skolnick et~al.}(1987)\citenamefont{Skolnick, Rorison,
  Nash, Mowbray, Tapster, Bass, and Pitt}}]{Skolnick1}
\bibinfo{author}{\bibfnamefont{M.~S.} \bibnamefont{Skolnick}},
  \bibinfo{author}{\bibfnamefont{J.~M.} \bibnamefont{Rorison}},
  \bibinfo{author}{\bibfnamefont{K.~J.} \bibnamefont{Nash}},
  \bibinfo{author}{\bibfnamefont{D.~J.} \bibnamefont{Mowbray}},
  \bibinfo{author}{\bibfnamefont{P.~R.} \bibnamefont{Tapster}},
  \bibinfo{author}{\bibfnamefont{S.~J.} \bibnamefont{Bass}}, \bibnamefont{and}
  \bibinfo{author}{\bibfnamefont{A.~D.} \bibnamefont{Pitt}},
  \bibinfo{journal}{Phys. Rev. Lett.} \textbf{\bibinfo{volume}{58}},
  \bibinfo{pages}{2130} (\bibinfo{year}{1987}).

\bibitem[{\citenamefont{Huard et~al.}(2000)\citenamefont{Huard, Cox,
  Saminadayar, Bourgognon, Arnoult, Cibert, and Tatarenko}}]{Hourd}
\bibinfo{author}{\bibfnamefont{V.}~\bibnamefont{Huard}},
  \bibinfo{author}{\bibfnamefont{R.}~\bibnamefont{Cox}},
  \bibinfo{author}{\bibfnamefont{K.}~\bibnamefont{Saminadayar}},
  \bibinfo{author}{\bibfnamefont{C.}~\bibnamefont{Bourgognon}},
  \bibinfo{author}{\bibfnamefont{A.}~\bibnamefont{Arnoult}},
  \bibinfo{author}{\bibfnamefont{J.}~\bibnamefont{Cibert}}, \bibnamefont{and}
  \bibinfo{author}{\bibfnamefont{S.}~\bibnamefont{Tatarenko}},
  \bibinfo{journal}{Physica E} \textbf{\bibinfo{volume}{6}},
  \bibinfo{pages}{161} (\bibinfo{year}{2000}).

\bibitem[{\citenamefont{Hawrylak}(1991)}]{Hawrylak}
\bibinfo{author}{\bibfnamefont{P.}~\bibnamefont{Hawrylak}},
  \bibinfo{journal}{Phys. Rev. B} \textbf{\bibinfo{volume}{44}},
  \bibinfo{pages}{3821} (\bibinfo{year}{1991}).

\bibitem[{\citenamefont{Teran et~al.}(2006)\citenamefont{Teran, Chen, Potemski,
  Wojtowicz, and Karczewski}}]{Teran}
\bibinfo{author}{\bibfnamefont{F.~J.} \bibnamefont{Teran}},
  \bibinfo{author}{\bibfnamefont{Y.}~\bibnamefont{Chen}},
  \bibinfo{author}{\bibfnamefont{M.}~\bibnamefont{Potemski}},
  \bibinfo{author}{\bibfnamefont{T.}~\bibnamefont{Wojtowicz}},
  \bibnamefont{and}
  \bibinfo{author}{\bibfnamefont{G.}~\bibnamefont{Karczewski}},
  \bibinfo{journal}{Phys. Rev. B} \textbf{\bibinfo{volume}{73}},
  \bibinfo{pages}{115336} (\bibinfo{year}{2006}).

\bibitem[{\citenamefont{Jusserand et~al.}(2001)\citenamefont{Jusserand,
  Karczewski, Cywi\'nski, Wojtowicz, Lema\^itre, Testelin, and
  Rigaux}}]{Jusserand1}
\bibinfo{author}{\bibfnamefont{B.}~\bibnamefont{Jusserand}},
  \bibinfo{author}{\bibfnamefont{G.}~\bibnamefont{Karczewski}},
  \bibinfo{author}{\bibfnamefont{G.}~\bibnamefont{Cywi\'nski}},
  \bibinfo{author}{\bibfnamefont{T.}~\bibnamefont{Wojtowicz}},
  \bibinfo{author}{\bibfnamefont{A.}~\bibnamefont{Lema\^itre}},
  \bibinfo{author}{\bibfnamefont{C.}~\bibnamefont{Testelin}}, \bibnamefont{and}
  \bibinfo{author}{\bibfnamefont{C.}~\bibnamefont{Rigaux}},
  \bibinfo{journal}{Phys. Rev. B} \textbf{\bibinfo{volume}{63}},
  \bibinfo{pages}{161302(R)} (\bibinfo{year}{2001}), \bibinfo{note}{and
  references therein}.

\bibitem[{\citenamefont{Jusserand et~al.}(2000)\citenamefont{Jusserand,
  Vijayaraghavan, Laruelle, Cavanna, and Etienne}}]{Jusserand2}
\bibinfo{author}{\bibfnamefont{B.}~\bibnamefont{Jusserand}},
  \bibinfo{author}{\bibfnamefont{M.~N.}~\bibnamefont{Vijayaraghavan}},
  \bibinfo{author}{\bibfnamefont{F.}~\bibnamefont{Laruelle}},
  \bibinfo{author}{\bibfnamefont{A.}~\bibnamefont{Cavanna}}, \bibnamefont{and}
  \bibinfo{author}{\bibfnamefont{B.}~\bibnamefont{Etienne}},
  \bibinfo{journal}{Phys. Rev. Lett.} \textbf{\bibinfo{volume}{85}},
  \bibinfo{pages}{5400} (\bibinfo{year}{2000}).

\bibitem[{\citenamefont{Pinczuk et~al.}(1979)\citenamefont{Pinczuk, Abstreiter,
  Trommer, and Cardona}}]{Pinczuk3}
\bibinfo{author}{\bibfnamefont{A.}~\bibnamefont{Pinczuk}},
  \bibinfo{author}{\bibfnamefont{G.}~\bibnamefont{Abstreiter}},
  \bibinfo{author}{\bibfnamefont{R.}~\bibnamefont{Trommer}}, \bibnamefont{and}
  \bibinfo{author}{\bibfnamefont{M.}~\bibnamefont{Cardona}},
  \bibinfo{journal}{Solid State commun.} \textbf{\bibinfo{volume}{30}},
  \bibinfo{pages}{429} (\bibinfo{year}{1979}).

\bibitem[{\citenamefont{Fasol et~al.}(1986)\citenamefont{Fasol, Mestres,
  Hughes, Fischer, and Ploog}}]{Fasol}
\bibinfo{author}{\bibfnamefont{G.}~\bibnamefont{Fasol}},
  \bibinfo{author}{\bibfnamefont{N.}~\bibnamefont{Mestres}},
  \bibinfo{author}{\bibfnamefont{H.~P.} \bibnamefont{Hughes}},
  \bibinfo{author}{\bibfnamefont{A.}~\bibnamefont{Fischer}}, \bibnamefont{and}
  \bibinfo{author}{\bibfnamefont{K.}~\bibnamefont{Ploog}},
  \bibinfo{journal}{Phys. Rev. Lett.} \textbf{\bibinfo{volume}{56}},
  \bibinfo{pages}{2517} (\bibinfo{year}{1986}).

\bibitem[{\citenamefont{Jusserand et~al.}(1992)\citenamefont{Jusserand,
  Richards, Peric, and Etienne}}]{Jusserand3}
\bibinfo{author}{\bibfnamefont{B.}~\bibnamefont{Jusserand}},
  \bibinfo{author}{\bibfnamefont{D.}~\bibnamefont{Richards}},
  \bibinfo{author}{\bibfnamefont{H.}~\bibnamefont{Peric}}, \bibnamefont{and}
  \bibinfo{author}{\bibfnamefont{B.}~\bibnamefont{Etienne}},
  \bibinfo{journal}{Phys. Rev. Lett.} \textbf{\bibinfo{volume}{69}},
  \bibinfo{pages}{848} (\bibinfo{year}{1992}).

\bibitem[{\citenamefont{Sarma and Wang}(1999)}]{Sarma}
\bibinfo{author}{\bibfnamefont{S.} \bibnamefont{Das~Sarma}} \bibnamefont{and}
  \bibinfo{author}{\bibfnamefont{D.-W.} \bibnamefont{Wang}},
  \bibinfo{journal}{Phys Rev. Lett.} \textbf{\bibinfo{volume}{83}},
  \bibinfo{pages}{816} (\bibinfo{year}{1999}).

\bibitem[{\citenamefont{Perez et~al.}(2006)\citenamefont{Perez, Jusserand,
  Richards, and Karczewski}}]{Perez1}
\bibinfo{author}{\bibfnamefont{F.}~\bibnamefont{Perez}},
  \bibinfo{author}{\bibfnamefont{B.}~\bibnamefont{Jusserand}},
  \bibinfo{author}{\bibfnamefont{D.}~\bibnamefont{Richards}}, \bibnamefont{and}
  \bibinfo{author}{\bibfnamefont{G.}~\bibnamefont{Karczewski}},
  \bibinfo{journal}{Physica Status Solidi B} \textbf{\bibinfo{volume}{243(4)}},
  \bibinfo{pages}{873} (\bibinfo{year}{2006}).

\bibitem[{\citenamefont{Pinczuk et~al.}(1992)\citenamefont{Pinczuk, Dennis,
  Heiman, Kallin, Brey, Tejedor, Schmitt-Rink, Pfeiffer, and West}}]{Pinczuk4}
\bibinfo{author}{\bibfnamefont{A.}~\bibnamefont{Pinczuk}},
  \bibinfo{author}{\bibfnamefont{B.~S.} \bibnamefont{Dennis}},
  \bibinfo{author}{\bibfnamefont{D.}~\bibnamefont{Heiman}},
  \bibinfo{author}{\bibfnamefont{C.}~\bibnamefont{Kallin}},
  \bibinfo{author}{\bibfnamefont{L.}~\bibnamefont{Brey}},
  \bibinfo{author}{\bibfnamefont{C.}~\bibnamefont{Tejedor}},
  \bibinfo{author}{\bibfnamefont{S.}~\bibnamefont{Schmitt-Rink}},
  \bibinfo{author}{\bibfnamefont{L.~N.} \bibnamefont{Pfeiffer}},
  \bibnamefont{and} \bibinfo{author}{\bibfnamefont{K.~W.} \bibnamefont{West}},
  \bibinfo{journal}{Phys. Rev. Lett.} \textbf{\bibinfo{volume}{68}},
  \bibinfo{pages}{3623} (\bibinfo{year}{1992}).

\bibitem[{cyn({\natexlab{a}})}]{cynt3}
\bibinfo{note}{The bare band-edge mass is the minimum conduction band effective
  mass defined as $m_b = \hbar^2 {[{\partial^2 E\over {\partial
  k^2}}]^{-1}}_{k=0}$, where $E$ is the band energy and $k$ is the wave vector.
  The effective mass $m^*_F = \hbar^2 [{dE\over dk}]^{-1}_{k_F}$.}

\bibitem[{\citenamefont{Landau and Lifshitz}(1962)}]{Landau}
\bibinfo{author}{\bibfnamefont{L.~D.} \bibnamefont{Landau}} \bibnamefont{and}
  \bibinfo{author}{\bibfnamefont{E.~M.} \bibnamefont{Lifshitz}},
  \emph{\bibinfo{title}{The Classical Theory of Fields}},
  \bibinfo{volume}{2nd Ed.} (\bibinfo{publisher}{Pergamon Press, Oxford},
  \bibinfo{year}{1962}).

\bibitem[{\citenamefont{Perez et~al.}(2007)\citenamefont{Perez, Aku-Leh,
  Richards, Jusserand, and Karczewski}}]{Florent2}
\bibinfo{author}{\bibfnamefont{F.}~\bibnamefont{Perez}},
  \bibinfo{author}{\bibfnamefont{C.}~\bibnamefont{Aku-Leh}},
  \bibinfo{author}{\bibfnamefont{D.}~\bibnamefont{Richards}},
  \bibinfo{author}{\bibfnamefont{B.}~\bibnamefont{Jusserand}},
  \bibnamefont{and}
  \bibinfo{author}{\bibfnamefont{G.}~\bibnamefont{Karczewski}},
  \bibinfo{journal}{Phys. Rev. Lett.~(to be published)}.

\bibitem[{\citenamefont{Lema\^itre et~al.}(2000)\citenamefont{Lema\^itre,
  Testelin, Rigaux, Wojtowicz, and Karczewski}}]{Lemaître}
\bibinfo{author}{\bibfnamefont{A.}~\bibnamefont{Lema\^itre}},
  \bibinfo{author}{\bibfnamefont{C.}~\bibnamefont{Testelin}},
  \bibinfo{author}{\bibfnamefont{C.}~\bibnamefont{Rigaux}},
  \bibinfo{author}{\bibfnamefont{T.}~\bibnamefont{Wojtowicz}},
  \bibnamefont{and}
  \bibinfo{author}{\bibfnamefont{G.}~\bibnamefont{Karczewski}},
  \bibinfo{journal}{Phys. Rev. B} \textbf{\bibinfo{volume}{62}},
  \bibinfo{pages}{5059} (\bibinfo{year}{2000}).

\bibitem[{\citenamefont{Astakhov and et~al.}(2002)}]{Astakhov}
\bibinfo{author}{\bibfnamefont{G.~V.} \bibnamefont{Astakhov}} \bibnamefont{and}
  \bibinfo{author}{\bibnamefont{et~al.}}, \bibinfo{journal}{Phys. Rev. B}
  \textbf{\bibinfo{volume}{65}}, \bibinfo{pages}{115310}
  (\bibinfo{year}{2002}).

\bibitem[{\citenamefont{Gubarev et~al.}(2000)\citenamefont{Gubarev, Kukushkin,
  Tovstonog, Akimov, Smet, von Klitzing, and Wegscheider}}]{Gubarev}
\bibinfo{author}{\bibfnamefont{S.~I.} \bibnamefont{Gubarev}},
  \bibinfo{author}{\bibfnamefont{I.~V.} \bibnamefont{Kukushkin}},
  \bibinfo{author}{\bibfnamefont{S.~V.} \bibnamefont{Tovstonog}},
  \bibinfo{author}{\bibfnamefont{M.~Y.} \bibnamefont{Akimov}},
  \bibinfo{author}{\bibfnamefont{J.}~\bibnamefont{Smet}},
  \bibinfo{author}{\bibfnamefont{K.}~\bibnamefont{von Klitzing}},
  \bibnamefont{and}
  \bibinfo{author}{\bibfnamefont{W.}~\bibnamefont{Wegscheider}},
  \bibinfo{journal}{JETP Lett.} \textbf{\bibinfo{volume}{72}},
  \bibinfo{pages}{324} (\bibinfo{year}{2000}).

\bibitem[{\citenamefont{Kukushkin and Timofeev}(1996)}]{Kukushkin}
\bibinfo{author}{\bibfnamefont{I.}~\bibnamefont{Kukushkin}} \bibnamefont{and}
  \bibinfo{author}{\bibfnamefont{V.~B.} \bibnamefont{Timofeev}},
  \bibinfo{journal}{Adv. Physics} \textbf{\bibinfo{volume}{45}},
  \bibinfo{pages}{147} (\bibinfo{year}{1996}).

\bibitem[{\citenamefont{Shields et~al.}(1996)\citenamefont{Shields, Osborne,
  Simmons, Ritchie, and Pepper}}]{Shields}
\bibinfo{author}{\bibfnamefont{A.~J.} \bibnamefont{Shields}},
  \bibinfo{author}{\bibfnamefont{J.~L.} \bibnamefont{Osborne}},
  \bibinfo{author}{\bibfnamefont{M.~Y.} \bibnamefont{Simmons}},
  \bibinfo{author}{\bibfnamefont{D.~A.} \bibnamefont{Ritchie}},
  \bibnamefont{and} \bibinfo{author}{\bibfnamefont{M.}~\bibnamefont{Pepper}},
  \bibinfo{journal}{Semicond. Sci. Technol.} \textbf{\bibinfo{volume}{11}},
  \bibinfo{pages}{890} (\bibinfo{year}{1996}).

\bibitem[{\citenamefont{Christen and Bimberg}(1990)}]{ChristenTheory}
\bibinfo{author}{\bibfnamefont{J.}~\bibnamefont{Christen}} \bibnamefont{and}
  \bibinfo{author}{\bibfnamefont{D.}~\bibnamefont{Bimberg}},
  \bibinfo{journal}{Phys. Rev. B} \textbf{\bibinfo{volume}{42}},
  \bibinfo{pages}{7213} (\bibinfo{year}{1990}).

\bibitem[{\citenamefont{Richards et~al.}(1990)\citenamefont{Richards, Fasol,
  and Ploog}}]{DavidPlasmons}
\bibinfo{author}{\bibfnamefont{D.}~\bibnamefont{Richards}},
  \bibinfo{author}{\bibfnamefont{G.}~\bibnamefont{Fasol}}, \bibnamefont{and}
  \bibinfo{author}{\bibfnamefont{K.}~\bibnamefont{Ploog}},
  \bibinfo{journal}{Appl. Phys. Lett.} \textbf{\bibinfo{volume}{56}},
  \bibinfo{pages}{1649} (\bibinfo{year}{1990}).

\bibitem[{cyn({\natexlab{b}})}]{cynt2}
\bibinfo{note}{The plasmon mode in our quantum wells needed too strong an
  illumination power to be excited, and coupled with the different resonance
  conditions with the SPEs, we therefore did not consider it.}

\bibitem[{\citenamefont{Matsuda et~al.}(2002)\citenamefont{Matsuda, Ikaida,
  Miura, Kuroda, Takano, and Takita}}]{Matsuda}
\bibinfo{author}{\bibfnamefont{Y.~H.} \bibnamefont{Matsuda}},
  \bibinfo{author}{\bibfnamefont{T.}~\bibnamefont{Ikaida}},
  \bibinfo{author}{\bibfnamefont{N.}~\bibnamefont{Miura}},
  \bibinfo{author}{\bibfnamefont{S.}~\bibnamefont{Kuroda}},
  \bibinfo{author}{\bibfnamefont{F.}~\bibnamefont{Takano}}, \bibnamefont{and}
  \bibinfo{author}{\bibfnamefont{K.}~\bibnamefont{Takita}},
  \bibinfo{journal}{Phys. Rev. B} \textbf{\bibinfo{volume}{65}},
  \bibinfo{pages}{115202} (\bibinfo{year}{2002}).

\bibitem[{\citenamefont{Smith and Stiles}(1972)}]{SmithStiles}
\bibinfo{author}{\bibfnamefont{J.~L.} \bibnamefont{Smith}} \bibnamefont{and}
  \bibinfo{author}{\bibfnamefont{P.~J.} \bibnamefont{Stiles}},
  \bibinfo{journal}{Phys. Rev. Lett.} \textbf{\bibinfo{volume}{29}},
  \bibinfo{pages}{102} (\bibinfo{year}{1972}).

\bibitem[{\citenamefont{Tang and et~al.}(2006)}]{Tang}
\bibinfo{author}{\bibfnamefont{N.}~\bibnamefont{Tang}} \bibnamefont{and}
  \bibinfo{author}{\bibnamefont{et~al.}}, \bibinfo{journal}{Appl. Phys. Lett.}
  \textbf{\bibinfo{volume}{88}}, \bibinfo{pages}{172115}
  (\bibinfo{year}{2006}).

\bibitem[{\citenamefont{Richards}(2000)}]{Richards}
\bibinfo{author}{\bibfnamefont{D.}~\bibnamefont{Richards}},
  \bibinfo{journal}{Phys. Rev. B} \textbf{\bibinfo{volume}{61}},
  \bibinfo{pages}{7517} (\bibinfo{year}{2000}).

\bibitem[{\citenamefont{Haufe et~al.}(1988)\citenamefont{Haufe, Schwabe,
  Fieseler, and Ilegems}}]{Haufe}
\bibinfo{author}{\bibfnamefont{A.}~\bibnamefont{Haufe}},
  \bibinfo{author}{\bibfnamefont{R.}~\bibnamefont{Schwabe}},
  \bibinfo{author}{\bibfnamefont{H.}~\bibnamefont{Fieseler}}, \bibnamefont{and}
  \bibinfo{author}{\bibfnamefont{M.}~\bibnamefont{Ilegems}},
  \bibinfo{journal}{J. Phys. C: Solid State Phys.}
  \textbf{\bibinfo{volume}{21}}, \bibinfo{pages}{2951} (\bibinfo{year}{1988}).

\bibitem[{\citenamefont{Perrin}(2006)}]{Perin}
\bibinfo{author}{\bibfnamefont{M.}~\bibnamefont{Perrin}}, Ph.D. thesis,
  \bibinfo{school}{Universit\'e Paris VI} (\bibinfo{year}{2006}).

\bibitem[{\citenamefont{Keller et~al.}(2005)\citenamefont{Keller, Yakovlev,
  Astakhov, Ossau, Crooker, Slobodskyy, Waag, Schmidt, and Molenkamp}}]{Keller}
\bibinfo{author}{\bibfnamefont{D.}~\bibnamefont{Keller}},
  \bibinfo{author}{\bibfnamefont{D.~R.} \bibnamefont{Yakovlev}},
  \bibinfo{author}{\bibfnamefont{G.~V.} \bibnamefont{Astakhov}},
  \bibinfo{author}{\bibfnamefont{W.}~\bibnamefont{Ossau}},
  \bibinfo{author}{\bibfnamefont{S.~A.} \bibnamefont{Crooker}},
  \bibinfo{author}{\bibfnamefont{T.}~\bibnamefont{Slobodskyy}},
  \bibinfo{author}{\bibfnamefont{A.}~\bibnamefont{Waag}},
  \bibinfo{author}{\bibfnamefont{G.}~\bibnamefont{Schmidt}}, \bibnamefont{and}
  \bibinfo{author}{\bibfnamefont{L.~W.} \bibnamefont{Molenkamp}},
  \bibinfo{journal}{Phys. Rev. B} \textbf{\bibinfo{volume}{72}},
  \bibinfo{pages}{235306} (\bibinfo{year}{2005}).

\bibitem[{\citenamefont{Nixon and Davies}(1990)}]{britguy}
\bibinfo{author}{\bibfnamefont{J.~A.} \bibnamefont{Nixon}} \bibnamefont{and}
  \bibinfo{author}{\bibfnamefont{J.~H.} \bibnamefont{Davies}},
  \bibinfo{journal}{Phys. Rev. B} \textbf{\bibinfo{volume}{41}},
  \bibinfo{pages}{7929} (\bibinfo{year}{1990}).

\bibitem[{\citenamefont{Peric}(1993)}]{PericTh}
\bibinfo{author}{\bibfnamefont{H.}~\bibnamefont{Peric}}, Ph.D. thesis,
  \bibinfo{school}{Universit\'e Paris XI} (\bibinfo{year}{1993}).

\bibitem[{\citenamefont{Kossut and Dobrowolski}(1993)}]{Kossut}
\bibinfo{author}{\bibfnamefont{J.}~\bibnamefont{Kossut}} \bibnamefont{and}
  \bibinfo{author}{\bibfnamefont{W.}~\bibnamefont{Dobrowolski}},
  \emph{\bibinfo{title}{Handbook of Magnetic Materials}},
  vol.~\bibinfo{volume}{7} (\bibinfo{publisher}{Elsevier},
  \bibinfo{year}{1993}).

\bibitem[{\citenamefont{Peyla et~al.}(1993)\citenamefont{Peyla, Wasiela, Merle~
  d'Aubign\'e, Ashenford, and Lunn}}]{Peyla}
\bibinfo{author}{\bibfnamefont{P.}~\bibnamefont{Peyla}},
  \bibinfo{author}{\bibfnamefont{A.}~\bibnamefont{Wasiela}},
  \bibinfo{author}{\bibfnamefont{Y.} \bibnamefont{Merle~d'Aubign\'e}},
  \bibinfo{author}{\bibfnamefont{D.~E.} \bibnamefont{Ashenford}},
  \bibnamefont{and} \bibinfo{author}{\bibfnamefont{B.}~\bibnamefont{Lunn}},
  \bibinfo{journal}{Phys. Rev. B} \textbf{\bibinfo{volume}{47}},
  \bibinfo{pages}{3783} (\bibinfo{year}{1993}).

\bibitem[{\citenamefont{Kohn}(1961)}]{Kohn}
\bibinfo{author}{\bibfnamefont{W.}~\bibnamefont{Kohn}}, \bibinfo{journal}{Phys.
  Rev.} \textbf{\bibinfo{volume}{123}}, \bibinfo{pages}{1242}
  (\bibinfo{year}{1961}).

\bibitem[{\citenamefont{Zhang and Sarma}(2006)}]{DasSarma2}
\bibinfo{author}{\bibfnamefont{Y.}~\bibnamefont{Zhang}} \bibnamefont{and}
  \bibinfo{author}{\bibfnamefont{S.} \bibnamefont{Das~Sarma}},
  \bibinfo{journal}{Phys. Rev. Lett.} \textbf{\bibinfo{volume}{96}},
  \bibinfo{pages}{196602} (\bibinfo{year}{2006}).

\bibitem[{\citenamefont{Attaccalite et~al.}(2002)\citenamefont{Attaccalite,
  Moroni, Gori-Giorgi, and Bachelet}}]{Attaccalite}
\bibinfo{author}{\bibfnamefont{C.}~\bibnamefont{Attaccalite}},
  \bibinfo{author}{\bibfnamefont{S.}~\bibnamefont{Moroni}},
  \bibinfo{author}{\bibfnamefont{P.}~\bibnamefont{Gori-Giorgi}},
  \bibnamefont{and} \bibinfo{author}{\bibfnamefont{G.~B.}
  \bibnamefont{Bachelet}}, \bibinfo{journal}{Phys. Rev. Lett.}
  \textbf{\bibinfo{volume}{88}}, \bibinfo{pages}{256601}
  (\bibinfo{year}{2002}).

\bibitem[{\citenamefont{Palo et~al.}(2005)\citenamefont{DePalo, Botti, Moroni,
  and Senatore}}]{DePalo}
\bibinfo{author}{\bibfnamefont{S.} \bibnamefont{DePalo}},
  \bibinfo{author}{\bibfnamefont{M.}~\bibnamefont{Botti}},
  \bibinfo{author}{\bibfnamefont{S.}~\bibnamefont{Moroni}}, \bibnamefont{and}
  \bibinfo{author}{\bibfnamefont{G.}~\bibnamefont{Senatore}},
  \bibinfo{journal}{Phys. Rev. Lett.} \textbf{\bibinfo{volume}{94}},
  \bibinfo{pages}{226405} (\bibinfo{year}{2005}).

\end{thebibliography}

\end{document}